# Electron-electron scattering processes in quantum wells in a quantizing magnetic field: II. Scattering in the case of two subbands*

M.P. Telenkov, Yu.A. Mityagin

P.N. Lebedev Physical Institute of the Russian Academy of Sciences, 119991, Moscow, Russia.

Electron–electron scattering processes involving Landau levels of two subbands are considered. A matrix of electron–electron scattering rates, containing all types of transitions between Landau levels, is calculated. This matrix is analyzed, and the relative rates of transitions of different types are determined. The effect of the quantizing magnetic field orientation on electron–electron scattering processes is established.

## 1. Introduction

In our previous work [1], we examined electron–electron scattering processes in quantum wells in a quantizing magnetic field. Expressions were obtained for the rate matrix of electron–electron scattering between Landau levels of different subbands in a quantizing magnetic field of arbitrary orientation relative to the structure layers. Calculations of the electron–electron scattering rates for transitions between Landau levels of a single quantum-well subband were performed, and trends in the behavior of the scattering matrix elements for all possible intrasubband transitions were revealed.

This work continues the research initiated in [1] on the analysis of the four-dimensional electron–electron scattering rate matrix, incorporating consideration of electron–electron scattering processes in the multi-subband case. An analysis of the influence of magnetic field orientation on electron–electron scattering processes in the Landau level system was also conducted.

## 2. Electron-electron scattering rate matrix in a quantizing magnetic field tilted with respect to the layers of a quantum well

To calculate the scattering rate matrix, we will utilize the model described in our previous work [1]. Here, we present only the basic expressions used directly in the calculations and necessary for a qualitative explanation of the obtained results.

In a magnetic field

$$\mathbf{B} = B_{\perp}\mathbf{e}_z + B_{\parallel}\mathbf{e}_y ,$$

where $z$ is the growth axis of a quantum well, energy levels, and wave functions of stationary states of an electron in a quantum well in the Landau gauge

$$\mathbf{A} = \left(B_{\parallel}z - B_{\perp}y\right)\mathbf{e}_x$$

are given by the expressions

$$E_{(\nu,n)} = \varepsilon_{\nu} + \Delta_{\nu}\left(B_{\parallel}\right) + \hbar\omega_{\perp}\left(n + \frac{1}{2}\right) \tag{1}$$

and

$$\psi(x,y,z) = \frac{\exp\left(ik_x x\right)}{\sqrt{L}}\varphi_{\nu}(z)\Phi_n\left(y - k_x\ell_{\perp}^2 - \langle z\rangle_{\nu}\, tg\theta\right). \tag{2}$$

Here $\varphi_{\nu}\left(z\right)$ is a wave function of the energy level $\varepsilon_{\nu}$ of the $\nu$ th subband, $\Phi_n\left(y\right)$ is a wave function of the n-th (n=0,1,2,…) energy level of a linear harmonic oscillator with a cyclotron frequency $\omega_{\perp} = eB_{\perp} / m_w c$, $m_w$ - effective mass in a quantum well, $\ell_{\perp} = \sqrt{\hbar / m_w\omega_{\perp}} = \sqrt{\hbar c / eB_{\perp}}$ - magnetic length,

$$\Delta_\nu\left(B_\parallel\right)=\frac{e^2}{2m_w c^2}\left(\delta z\right)_\nu^2\cdot B_\parallel^2, \tag{3}$$

$$\left\langle z\right\rangle_\nu=\int dz\,\overset{*}{\varphi}(z)z\varphi\left(z\right) \tag{4}$$

is the mean value of the electron z coordinate, $\left(\delta z\right)_\nu$ - its standard deviation, $\theta$ - magnetic field tilt angle ($tg\theta=B_\parallel/B_\perp$), $L$ – transverse size of a heterostructure.

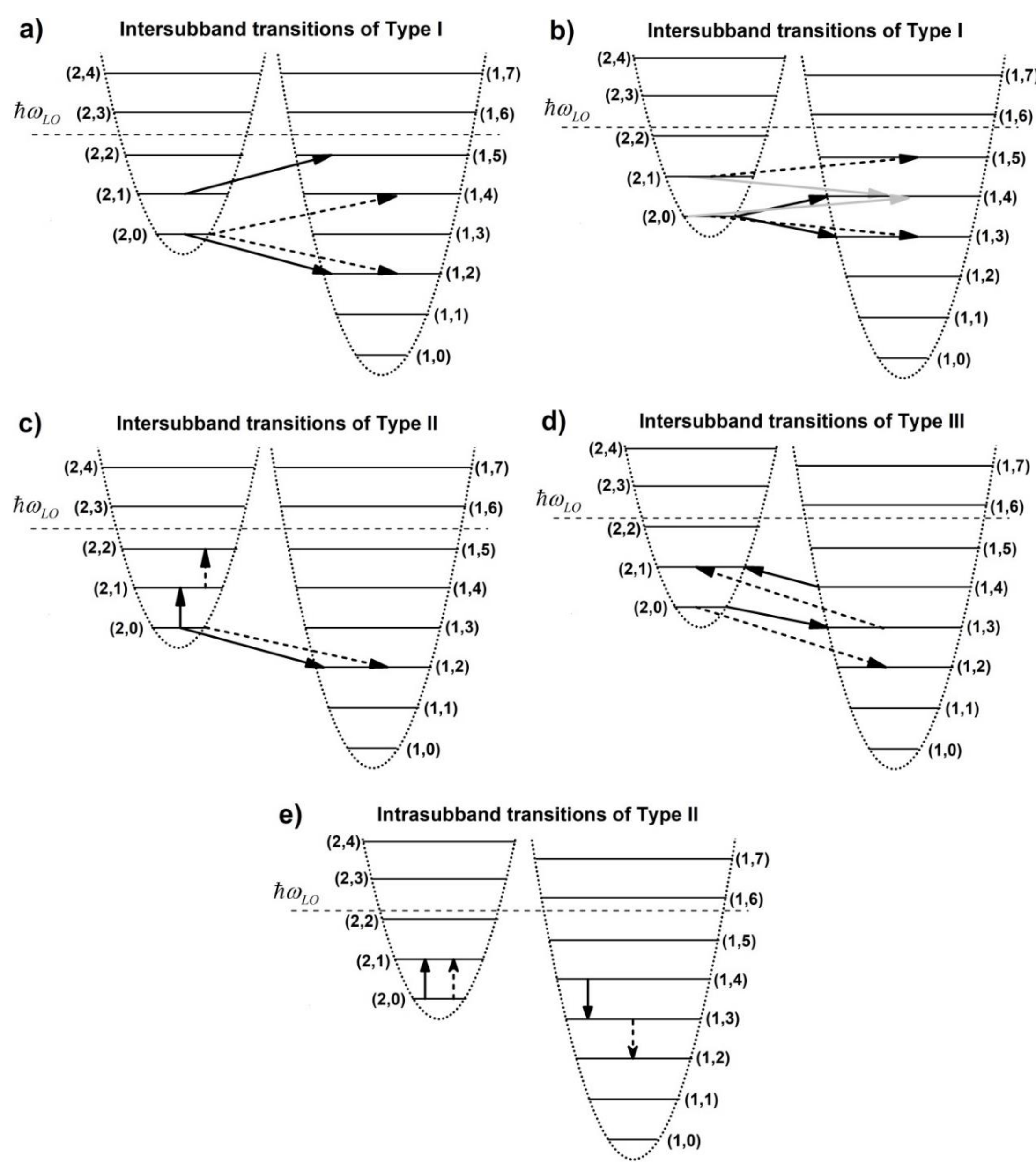

Figure 1. Different types of transitions involving Landau levels of different subbands. Arrows of different types indicate different transitions of the first and second electrons of the interacting pair.

Each Landau level is degenerate in the projection of the wave vector $k_x$ with multiplicity

$$\alpha = \frac{e}{\pi\hbar c}\cdot B_{\perp}. \qquad (5)$$

In what follows, the Landau level will be denoted by $i = (\nu_i, n_i)$, where $n_i$ is the number of the Landau level in the $\nu_i$-th subband.

The transition due to electron-electron interaction, in which one electron moves from the Landau level *i* to the Landau level *f*, and another electron moves from the Landau level *j* to the Landau level *g*, will be denoted by $\{i \to f; j \to g\}$.

The scattering rate matrix determines the electron fluxes between Landau levels. The electron flux from a Landau level *i* (the average number of electrons leaving the level per unit time, per unit area of the heterostructure)

$$J_{e-e}^{i,-} = \sum_{j,f,g} J_{e-e}\begin{pmatrix} i & j \\ f & g \end{pmatrix}, \qquad (6)$$

where

$$J\begin{pmatrix} i & j \\ f & g \end{pmatrix} = W_{e-e}\begin{pmatrix} i & j \\ f & g \end{pmatrix}\cdot N_i N_j \left[1-\frac{N_g}{\alpha}\right]\left[1-\frac{N_f}{\alpha}\right], \qquad (7)$$

is the total intensity of electron scattering from Landau levels i and j to Landau levels f and g, respectively (transition $\{(\nu_i, n_i) \to (\nu_f, n_f) \,\&(\nu_j, n_j) \to (\nu_g, n_g)\}$), $N_i$ is the number of electrons at Landau level *i*, per unit area of the structure (Landau level population).

The scattering rate matrix element for a given transition has the form

$$W_{e-e}\begin{pmatrix} i & j \\ f & g \end{pmatrix} = A_{e-e}\begin{pmatrix} i & j \\ f & g \end{pmatrix} F_{ee}\left(E_i + E_j - E_f - E_j\right), \qquad (8)$$

where the transition amplitude is given by the expression

$$A_{e-e}\begin{pmatrix} i & j \\ f & g \end{pmatrix} = \frac{4e^4}{\pi^2\varepsilon_s^2\hbar}\,\frac{\exp\left(\left[-\dfrac{(\xi_{\nu_f,\nu_i}+\xi_{\nu_g,\nu_j})^2}{4}\right]\right)}{2^{n_i+n_j+n_g+n_f}\, n_i!\,n_j!\,n_g!\,n_f!}\times$$
$$\times\iint dk_1 dk_2 \exp\left(-\left[k_2 - \frac{(\xi_{\nu_i,\nu_j}+\xi_{\nu_g,\nu_f})}{2}\right]^2\right)\left|M_{i,j,g,f}\left(k_1,k_2\right)\right|^2, \qquad (9)$$

$$
\begin{aligned}
M_{i,j,g,f}\left(k_1,k_2\right) &= \int dy \exp\left(-\frac{1}{2}\left(y+k_1-k_2\right)^2\right)\times \\
&\times G_{\nu_i,\nu_j,\nu_g,\nu_f}\left(\left|k_2\right|;\left|y+\frac{\xi_{\nu_i,\nu_j}+\xi_{\nu_f,\nu_g}}{2}\right|\right)\Lambda_{n_i,n_j,n_g,n_f}\left(k_2,y+k_1-k_2\right)
\end{aligned}
\tag{10}
$$

$$
G_{\nu_i,\nu_j,\nu_g,\nu_f}(\gamma;y) = \int dz_1 K_0\left(\gamma\sqrt{y^2+\frac{4z_1^2}{\ell_\perp^2}}\right)\times R_{\nu_i,\nu_j,\nu_g,\nu_f}(z_1),
\tag{11}
$$

$$
R_{\nu_i,\nu_j,\nu_g,\nu_f}(z_1) = \int dz_2 \varphi_{\nu_i}(z_2)\varphi_{\nu_j}(z_2-2z_1)\overset{*}{\varphi}_{\nu_g}(z_2-2z_1)\overset{*}{\varphi}_{\nu_f}(z_2),
\tag{12}
$$

$$
\begin{aligned}
\Lambda_{n_i,n_j,n_g,n_f}(k,y) &= 2^{\frac{n_i+n_j+n_g+n_f}{2}}\sqrt{2}\sum_{p_1=0}^{n_i}\frac{1}{2^{p_1/2}}\binom{n_i}{p_1}H_{p_1}(\delta_i)\sum_{p_2=0}^{n_j}\frac{1}{2^{p_2/2}}\binom{n_j}{p_2}H_{p_2}(\delta_j)\times \\
&\times\sum_{p_3=0}^{n_g}\frac{1}{2^{p_3/2}}\binom{n_g}{p_3}H_{p_3}(\delta_g)\times\sum_{p_4=0}^{n_f}\frac{1}{2^{p_4/2}}\binom{n_f}{p_4}H_{p_4}(\delta_f)\times \\
&\times\frac{\left[1+\left(-1\right)^{n_i+n_j+n_g+n_f-p_1-p_2-p_3-p_4}\right]}{2}\times \\
&\times\Gamma\left(\frac{n_i+n_j+n_g+n_f-p_1-p_2-p_3-p_4}{2}+\frac{1}{2}\right)
\end{aligned}
,\tag{13}
$$

$K_0(x)$ - McDonald function, $\Gamma(x)$ - Euler gamma function, $H_n(x)$ - Hermite polynomial of degree n,

$$
\delta_i = \frac{y+k+\xi_{\nu_f,\nu_i}}{2},
\tag{14}
$$

$$
\delta_j = \frac{-y-k+\xi_{\nu_g,\nu_j}}{2},
\tag{15}
$$

$$
\delta_g = \frac{-y+k-\xi_{\nu_g,\nu_j}}{2},
\tag{16}
$$

$$
\delta_f = \frac{y-k-\xi_{\nu_f,\nu_i}}{2}.
\tag{17}
$$

$$
\xi_{\nu_1,\nu_2} = \sqrt{\frac{e}{\hbar c B_\perp}}\cdot\left[\left\langle z\right\rangle_{\nu_1}-\left\langle z\right\rangle_{\nu_2}\right]\cdot B_\parallel,
\tag{18}
$$

The form factor expressing the energy conservation law in electron-electron scattering, and taking into account the finite width of the Landau levels, is equal to

$$
F_{ee}\left(E_i+E_j-E_f-E_j\right) = \frac{1}{\sqrt{2\pi}\Gamma_\Sigma^{ee}}\exp\left\{-\frac{\left(E_i+E_j-E_f-E_j\right)^2}{2\left(\Gamma_\Sigma^{ee}\right)^2}\right\}.
\tag{19}
$$

The obtained regularities are illustrated using the example of a GaAs/$Al_{0.3}Ga_{0.7}As$ quantum well with a width of 25 nm, unless otherwise specified. In such a quantum well, two subbands lie below the optical phonon level, and electron–electron scattering is the primary mechanism of intersubband relaxation [2]. For such quantum wells, the typical Landau level width is 1 meV, which yields a transition width in form factor (18) of $\Gamma_{\Sigma}^{ee}$ approximately 2 meV.

Since all the main features of the electronic spectrum in the system under consideration, which principally distinguish it from other electronic systems, are determined by the magnetic field component directed along the growth axis of the heterostructure, to highlight the role of these features most clearly, we first consider the scattering processes in a magnetic field directed perpendicular to the heterostructure layers.

## 3. Intrasubband transitions caused by the interaction of electrons in different subbands

In the multi-subband case, a special type of transition arises: each electron of the interacting pair remains after scattering in the same subband in which it was initially located, but, unlike the intrasubband transitions considered in [1], these electrons are located in different subbands (Fig. 1e). We will call such transitions intrasubband transitions of type II, since for each electron such a transition is intrasubband. Accordingly, the transitions previously considered in [1], when the initial and final Landau levels of both electrons are located within the same subband ($\nu_i = \nu_j = \nu_f = \nu_g$), will be referred to as intrasubband transitions of type I.

Electron–electron scattering is an elastic two-particle process. For the transition

$$\left\{(\nu_i, n_i) \to (\nu_f, n_f); (\nu_j, n_j) \to (\nu_g, n_g)\right\}$$

the energy conservation law (the transition resonance condition) has the form

$$E_{(\nu_i, n_i)} + E_{(\nu_j, n_j)} = E_{(\nu_f, n_f)} + E_{(\nu_g, n_g)}. \tag{20}$$

Substituting expression (1) for single-particle energy into (20) yields

$$n_f - n_i + n_g - n_j = \frac{\left(\varepsilon_{\nu_i} - \varepsilon_{\nu_f} + \varepsilon_{\nu_j} - \varepsilon_{\nu_g}\right)}{\hbar\omega_c}. \tag{21}$$

For intrasubband transitions of type II, $\nu_i = \nu_f, \nu_j = \nu_g$, therefore, the resonance condition (21) takes the form

$$n_f - n_i + n_g - n_j = 0. \tag{22}$$

and, accordingly, is satisfied at any magnetic field value. As a result, the nature of the dependence of the rate of type II transitions on the magnetic field is the same as for type I intrasubband transitions: the transition rate slowly decreases with increasing magnetic field strength (Fig. 2). In this case, the change in the Landau level number of one electron in electron–electron scattering is equal in magnitude and opposite in sign to the change in the Landau level number of another electron. Therefore, the same types of transitions arise here as for type I, with the only difference being that the electrons move along a ladder of Landau levels in different subbands. The set of all type II intrasubband transitions for which the law of conservation of energy is satisfied can be described by the formulas

$$\left\{(\nu_i,n_i)\to(\nu_i,n_i-\Delta n);(\nu_j,n_j)\to(\nu_j,n_j+\Delta n)\right\} \qquad (23)$$

and

$$\left\{(\nu_i,n_i)\to(\nu_i,n_i+\Delta n),(\nu_j,n_j)\to(\nu_j,n_j-\Delta n)\right\}, \qquad (24)$$

Here, as in the case of type I transitions, the energy transferred to the electron is proportional to the number of levels through which the electron "jumps".

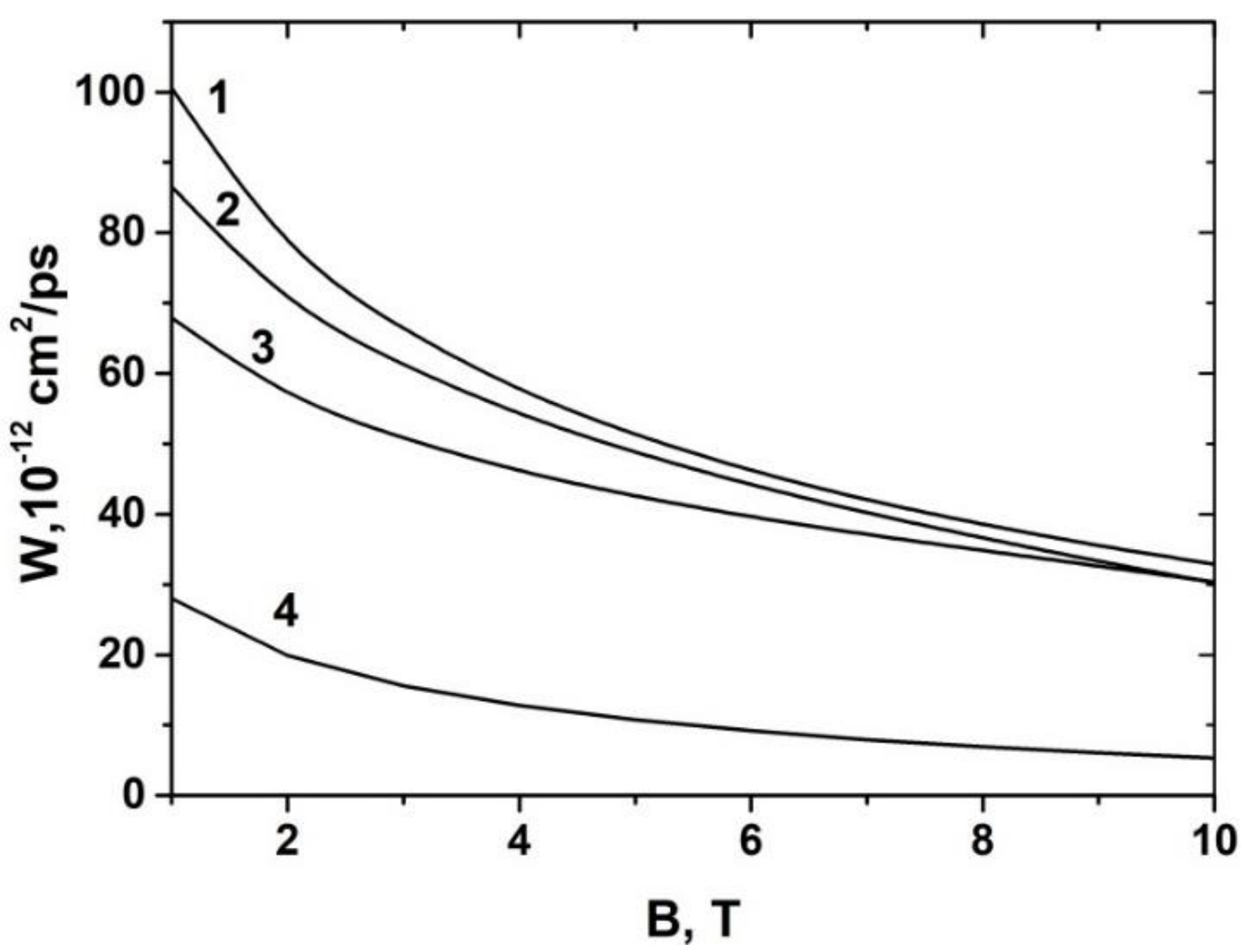

Figure 2. Dependence of the electron-electron scattering rate on the magnetic field for intrasubband transitions of type II: 1- $\left\{(2,0)\to(2,1)\,\&\,(1,1)\to(1,0)\right\}$;

2 - $\left\{(2,1)\to(2,2)\,\&\,(1,1)\to(1,0)\right\}$; 3- $\left\{(2,0)\to(2,1)\,\&\,(1,4)\to(1,3)\right\}$;

4 - $\left\{(2,0)\to(2,2)\,\&\,(1,2)\to(1,0)\right\}$.

For type II transitions, the dependence on the energy transferred to the electron is similar to that for type I transitions. The rate of type II transitions decreases sharply with increasing transferred energy (Fig. 3). The explanation is the same as in the case of type I intrasubband transitions (see [1]).

It's important to note that the rates of type I and type II transitions are pretty close in magnitude (refer to Fig. 4). This similarity is critical for understanding how electrons relax from the Landau levels of the upper subband, which lies below the energy of the optical phonon. Type II transitions create and activate a strong relaxation channel when electrons are excited into the upper subband. In this process, electrons move along the ladder of Landau levels within the upper subband and transit to the lower subband due to intersubband electron-phonon scattering processes [2].

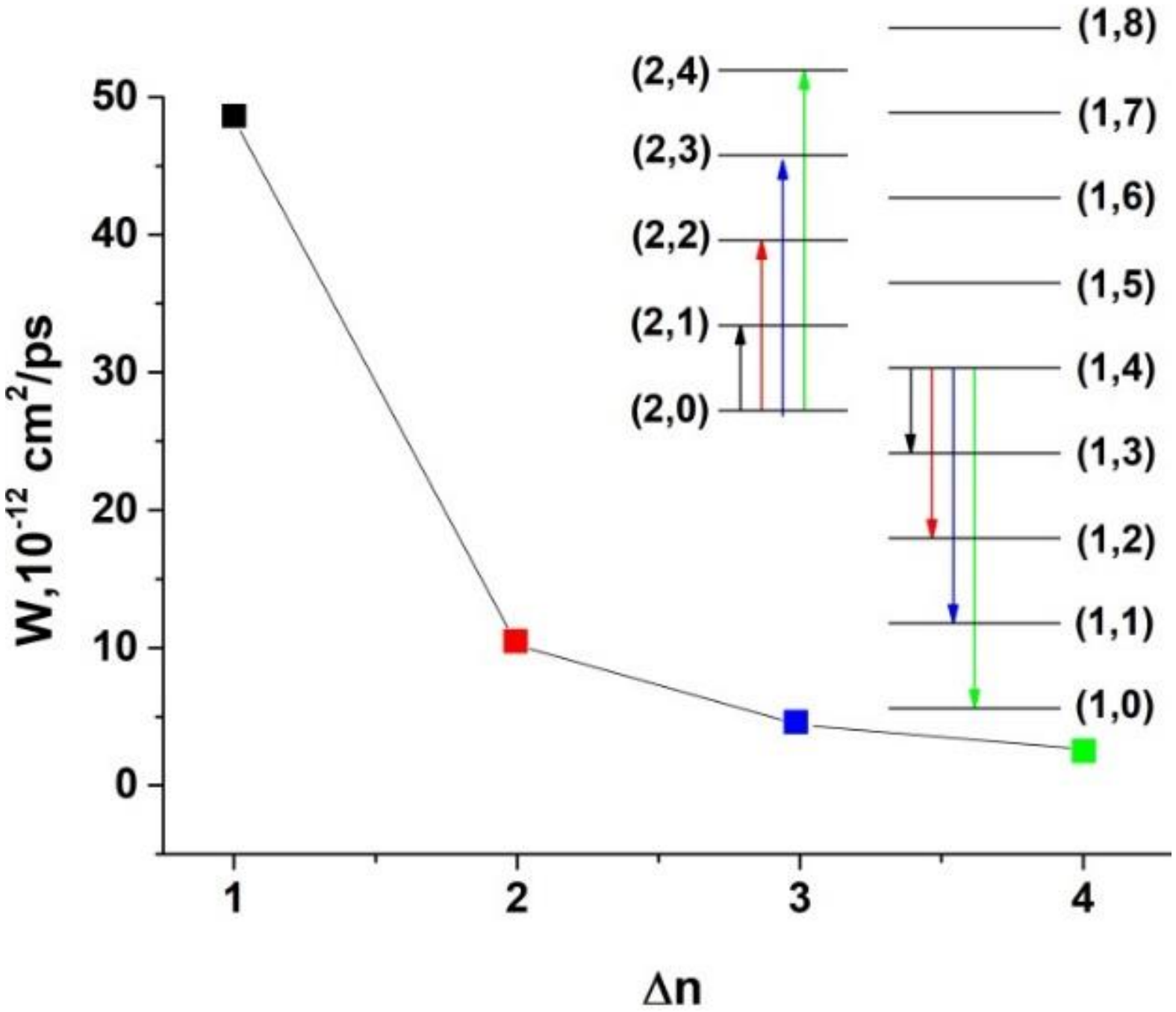

Figure 3. Dependence of the electron-electron scattering rate for type II intrasubband transitions on the change in Δn of the Landau level number of the electron at the transition. Magnetic field B=3.5 T.

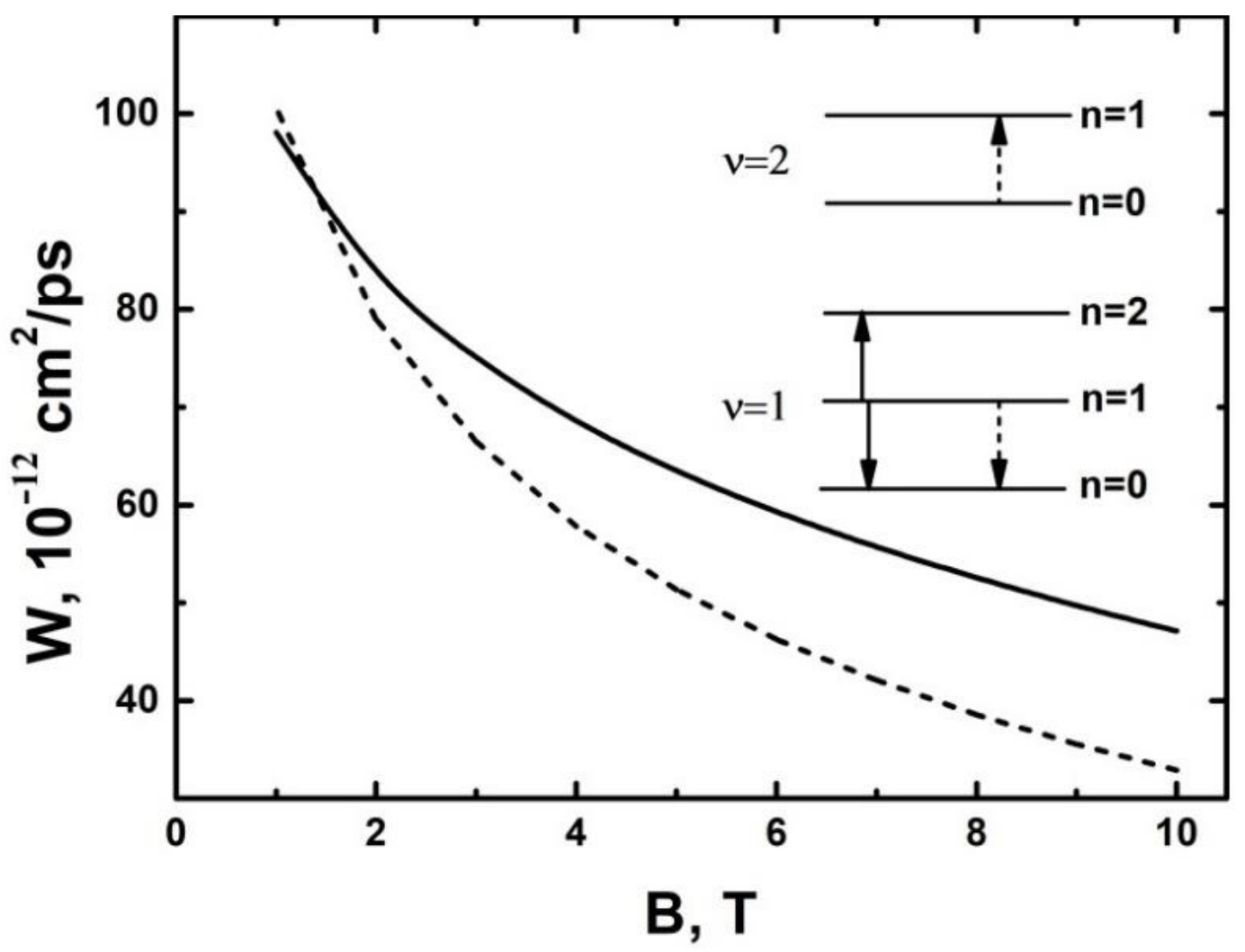

Figure 4. Dependence of the electron-electron scattering rate on the magnetic field for the intrasubband transition of type I $\{(1,1)\rightarrow(1,0)\,\&\,(1,1)\rightarrow(1,2)\}$ (solid curve) and intrasubband transition of type II $\{(1,1)\rightarrow(1,0)\,\&\,(2,0)\rightarrow(2,1)\}$ (dashed line).

Figure 5 illustrates the importance of type II transitions in the intersubband relaxation processes. It depicts the temporal dependence of the electron subsystem energy when selectively excited to the Landau level (2,0), which is positioned below the optical phonon energy. In the first curve (curve 1), all electron-electron scattering processes from Landau levels below the optical phonon energy are taken into account. In the second curve (curve 2), intrasubband transitions of type II are disregarded. It is evident that neglecting type II transitions significantly slows down the relaxation rate, resulting in a relaxation time that is more than three times longer.

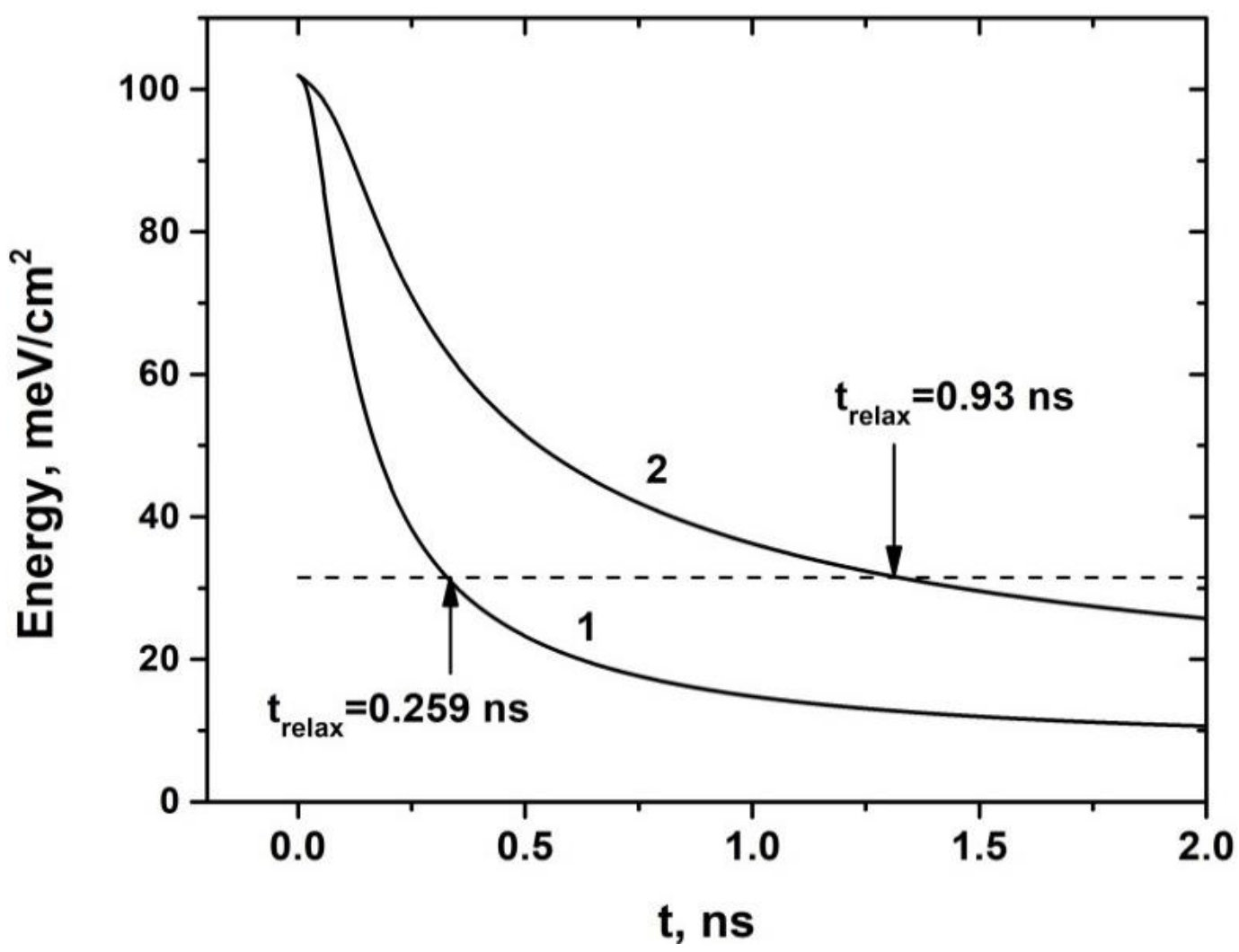

Fig. 5. Time dependence of the excitation energy of the electron subsystem during its selective excitation to the Landau level (2,0). Magnetic field B = 3.5 T (at this value, the optical phonon level passes near the Landau level (2,3)). Crystal lattice temperature $T_L$ = 4.2 K. The electron concentration in the well is 1.5 x $10^{10}$ $cm^{-2}$. Nonequilibrium is created by the instantaneous excitation of electrons to the Landau level (2,0), such that at the initial instant t = 0, only two Landau levels are populated: the (1,0) level and the (2,0) level, with concentrations of $10^{10}$ $cm^{-2}$ and 5 x $10^9$ $cm^{-2}$, respectively. Curve 1 is obtained by taking into account all transitions from Landau levels lying below the optical phonon energy. In obtaining curve 2, intrasubband transitions of type II are neglected.

### 5. Intersubband transitions

We divide the electron-electron scattering transitions between two subbands into three types.

In intersubband transitions of type I, both electrons of the interacting pair are in one subband ($\nu_i = \nu_j$) and, as a result of scattering, move to the Landau levels of the other subband ($\nu_g = \nu_f \neq \nu_i = \nu_j$). Examples of transitions of this type are shown in Fig. 2a and Fig. 2b.

In type II, we refer to transitions in which only one electron moves to the other subband while the second electron remains in the initial subband, i.e. $\nu_j \neq \nu_g$. Examples of transitions of this type are shown in Fig. 2c.

In the initial state of transitions of type III, the electrons of the interacting pair are in different subbands ($\nu_i \neq \nu_j$), and each of them transits to the Landau levels of the other subband ($\nu_f = \nu_j$ and $\nu_g = \nu_i$). Examples of this type of transition are shown in Fig. 2d.

In contrast to intrasubband scattering, the nature of the rate dependence on the magnetic field of intersubband transitions depends on their type.

In transitions of type I, the energy conservation law $E_{(\nu_i,n_i)} + E_{(\nu_i,n_j)} = E_{(\nu_f,n_f)} + E_{(\nu_f,n_g)}$ takes the form

$$n_f + n_g - n_i - n_j = 2\frac{\Delta\varepsilon_{if}}{\hbar\omega_\perp}, \tag{25}$$

where

$$\Delta\varepsilon_{if} = \varepsilon_{\nu_i} - \varepsilon_{\nu_f} \tag{26}$$

is the intersubband spacing.

This condition is satisfied in two cases: if $\frac{\Delta\varepsilon_{if}}{\hbar\omega_\perp}$ is an integer

$$\frac{\Delta\varepsilon_{if}}{\hbar\omega_\perp} = p, \tag{27}$$

where $p = 1,2,3\ldots$;

or $\frac{\Delta\varepsilon_{if}}{\hbar\omega_\perp}$ is a half-integer

$$\frac{\Delta\varepsilon}{\hbar\omega_\perp} = p + \frac{1}{2}, \tag{28}$$

where $p = 0,1,2,3\ldots$.

Condition (44) corresponds to the value of the magnetic field

$$B_{\perp,n} = \frac{m_w c}{\hbar e}\frac{\Delta\varepsilon_{if}}{p}, \qquad (29)$$

at which the 0-th Landau level of the upper subband coincides with the p-th Landau level of the lower subband ( $E_{(\nu_i,0)} = E_{(\nu_f,p)}$). Since the system of Landau levels of a subband is equidistant and the distance between neighboring levels is the same in each subband, the alignment with the Landau levels of the lower subband will simultaneously occur for all Landau levels of the upper subband – $E_{(\nu_i,n)} = E_{(\nu_f,p+n)}$, where $n = 0,1,2,\ldots$ (see Fig. 6). Therefore, at the value of the magnetic field satisfying condition (46), a large number of transitions are simultaneously in resonance. All these transitions can be described by the formula

$$\left\{(\nu_i,n)\rightarrow\left(\nu_f,p+n-\delta n\right)\&\left(\nu_i,n+\Delta N\right)\rightarrow\left(\nu_f,p+n+\Delta N+\delta n\right)\right\}. \qquad (30)$$

In this formula, n is the number of the initial Landau level of the electron in the second subband whose energy decreases in the scattering act. The energy transferred to one electron in the scattering act is

$$E_{trans} = \left|E_f - E_i\right| = \left|E_g - E_j\right| = \hbar\omega_c \cdot \delta n .$$

The difference between the energy in the initial state of the electron that moves to the higher level and the energy of the second electron of the interacting pair is $\Delta E_{init} = \hbar\omega_c \cdot \Delta N$.

In formula (30) $\delta n$ is the energy transferred to one electron in the scattering act, expressed in units of $\hbar\omega_c$. However, $\delta n$ now differs from the change in the Landau level number in the subband at the transition, as for an electron whose energy decreases with scattering,

$$n_f - n_i = p - \delta n , \qquad (31)$$

and for the electron whose energy increases during scattering,

$$n_g - n_j = p + \delta n . \qquad (32)$$

Depending on the sign of the difference $\Delta N$ between the numbers of the initial Landau levels, we have different types (subtypes) of type I intersubband transitions.

In case of $\Delta N = 0$ formula (30) describes transitions, when both electrons are at the same Landau level $\left(\nu_i,n\right)$ in the initial state. In this case $\Delta n \le p + n$.

The value $\Delta N \ne 0$ corresponds to the transitions in which the electrons of the interacting pair are initially at different Landau levels - the electron whose energy decreases in a scattering event is at the level $(\nu_i, n)$, and the second electron is at the level $(\nu_i, n+\Delta N)$

.

An example of transitions when $\Delta N > 0$ is shown in Figure 6a. In this case $\Delta n \le p+n$. An example of transitions when $\Delta N < 0$ is shown in Figure 6b. In this case $\Delta N \ge -n$.

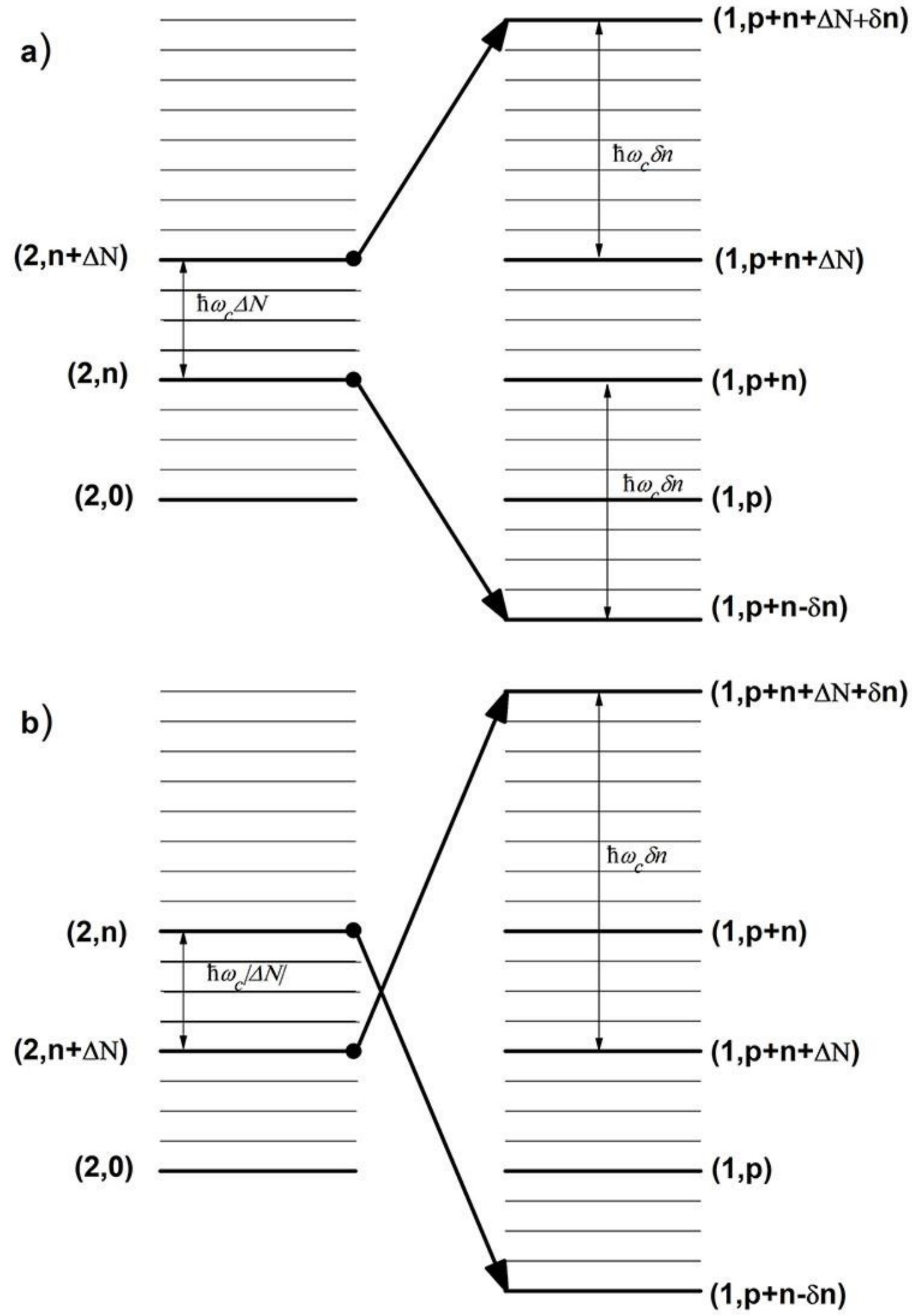

Figure 6. Scheme of type I intersubband transitions in the case when Landau levels of the upper subband coincide with Landau levels of the lower subband: a) for $\Delta N \ge 0$; b) for $\Delta N < 0$.

Condition (28) corresponds to the value of the magnetic field

$$B_{\perp,n} = \frac{m_w c}{\hbar e} \frac{\Delta\varepsilon_{if}}{p+\frac{1}{2}}, \tag{33}$$

at which the 0-th Landau level of the upper subband lies precisely in the middle between the p-th and (p+1)-th Landau levels of the lower subband

$(E_{(\nu_i,0)} = (E_{(\nu_f,p)} + E_{(\nu_f,p+1)})/2)$.

Correspondingly, all higher levels of the upper subband lie exactly midway between the neighboring Landau levels of the lower subband

$$E_{(\nu_i,n)} = (E_{(\nu_f,p+n)} + E_{(\nu_f,p+n+1)})/2,$$

where n=0,1,2… (see Fig. that7a). Thus, the resonance condition is simultaneously satisfied for a set of transitions that can be described by the formula

$$\left\{(\nu_i,n) \to \left(\nu_f, p+n+\frac{1}{2}-\delta n\right) \& (\nu_i, n+\Delta N) \to \left(\nu_f, p+n+\Delta N+\frac{1}{2}+\delta n\right)\right\}. \qquad (34)$$

Also, as before, $\delta n$ is the energy transferred to one electron expressed in units of $\hbar\omega_c$, i.e. $E_{trans} = \left|E_f - E_i\right| = \left|E_g - E_j\right| = \hbar\omega_c \cdot \delta n$, but now $\delta n$ takes half-integer positive values, less than or equal to $p+n+1/2$. At $\delta n = 1/2$, the electrons pass to the nearest Landau levels of the lower subband, at $\delta n = 3/2$, to the levels of the lower subband following the nearest ones, etc. In formula (51), $\Delta N$ can take integer values greater than $-\delta n$. Schemes of transitions with $\Delta N \geq 0$ and $\Delta N < 0$ are shown in Figures 17a and 17b, respectively. The change of the Landau level number in the subband for an electron whose energy decreases as a result of the interaction is

$$n_f - n_i = p - \left(\delta n - \frac{1}{2}\right), \qquad (35)$$

and for the electron whose energy increases

$$n_g - n_j = p + \left(\delta n + \frac{1}{2}\right). \qquad (36)$$

For each of type I intersubband transition, the energy conservation law is satisfied at only one specific value of the magnetic field. This value is determined either by the expression in (29) or by the one in (33). Given that Landau levels have a finite width, the dependence of the type I intersubband transition rate on the magnetic field manifests as a resonance peak (see Fig. 8).

A comparison of intrasubband scattering rates indicates that the rates of type I intersubband scattering—even at resonance—are significantly lower (by about two orders of magnitude) than those for both type I and type II intrasubband scattering. However, type I intersubband transitions facilitate the transfer of electrons from the upper subband to the lower subband, making them an essential channel for intersubband relaxation.

For type I intersubband transitions, it is interesting to ask whether it is possible to make a simple separation of matrix elements by their magnitude, similar to how we do for intrasubband transitions by the magnitude of energy transferred by scattering to one electron. At first glance, it would seem that for intersubband transitions of type I, there should be a dependence of the transition rate on $E_{trans}$, similar to intrasubband transitions – the rate of intersubband transitions of type I should decrease with an increase $E_{trans}$. However, the scattering rate matrix analysis has revealed this is not the case.

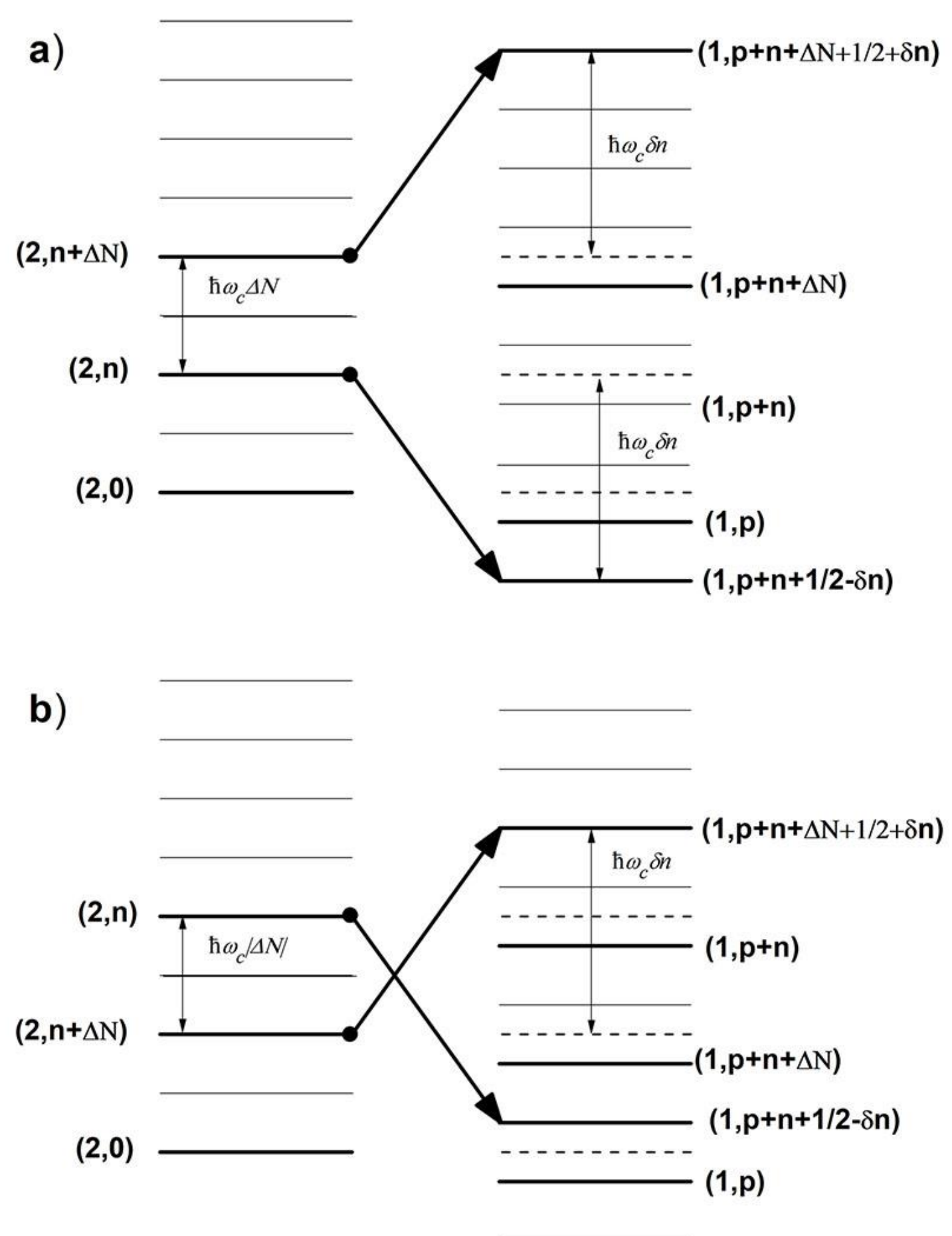

Figure 7. The scheme of type I intersubband transitions when the Landau levels of the upper subband are in the middle between the Landau levels of the lower subband; a) for $\Delta N > 0$; б) for $\Delta N < 0$.

Figure 9 shows the dependencies of the rate of type I intersubband transitions on the transferred energy. In each of the transitions presented, the initial states of the electrons in the interacting pair are fixed (n and its own $\Delta N$ for each figure), and the final states change as the energy (number $\delta n$) transferred to one electron in the scattering act increases, which is plotted on the abscissa axis. For intersubband transitions, there is no such clear dependence of the scattering rate on the transferred energy as in the case of intrasubband transitions. In the case when both electrons scatter from the (2,0) level, the scattering rate decreases rather rapidly with the transferred energy (Fig. 9a). At the same time, when both electrons scatter from different Landau levels, the rates of the first few transitions can be close in magnitude (Fig. 9b), and even exceed the transition with minimal change in electron energy (Fig. 9c). A rather complex non-monotonic dependence on the transferred energy is also possible (Fig. 9d).

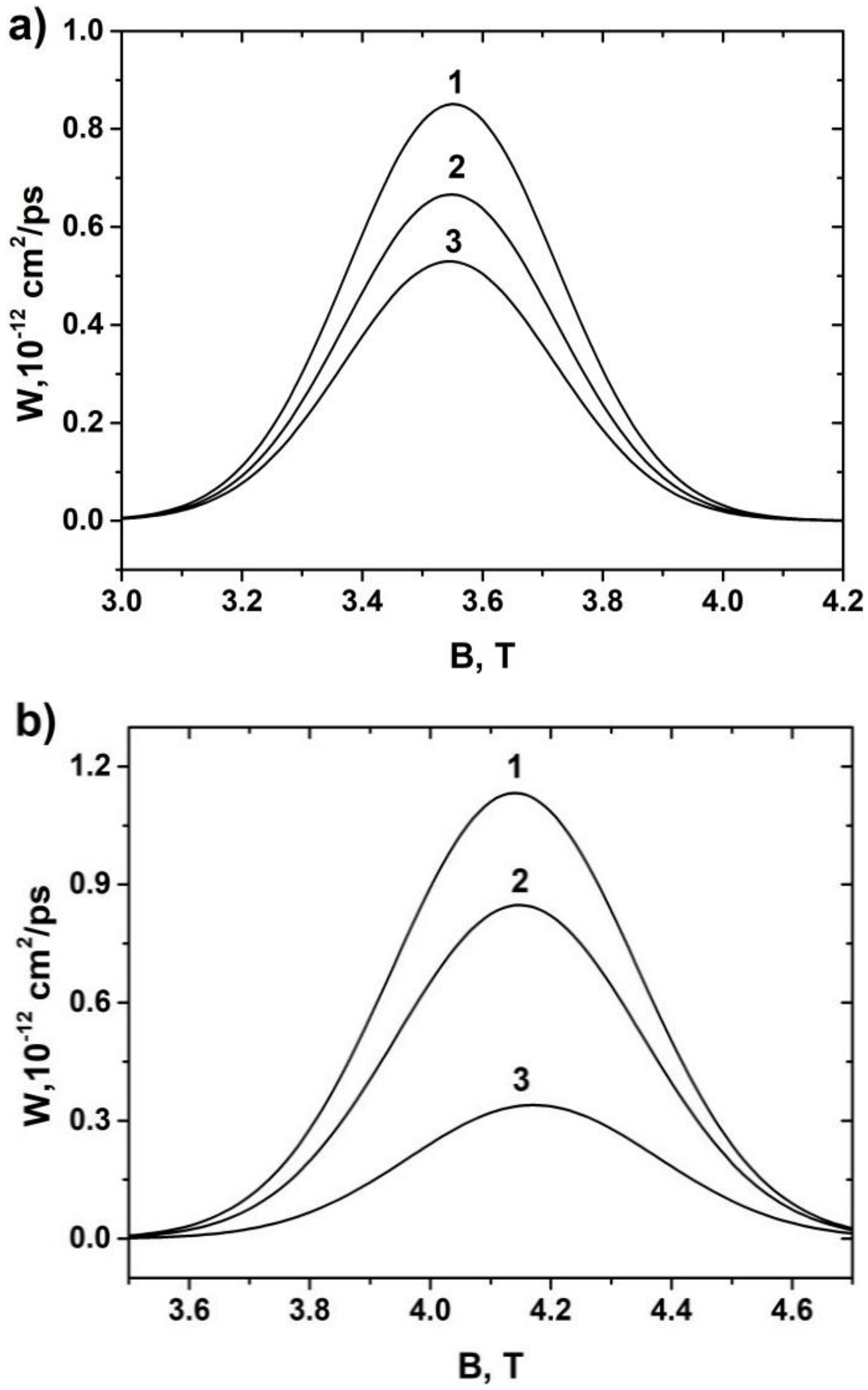

Figure 18. Dependence of the electron-electron scattering rate on the magnetic field for transitions of the type I.

a) 1 - $\{(2,0)\to(1,3)\,\&\,(2,0)\to(1,4)\}$, 2- $\{(2,0)\to(1,3)\,\&\,(2,1)\to(1,5)\}$,

3- $\{(2,0)\to(1,3)\,\&\,(2,2)\to(1,6)\}$.

б) 1- $\{(2,0)\to(1,3)\,\&\,(2,0)\to(1,3)\}$, 2- $\{(2,0)\to(1,2)\,\&\,(2,0)\to(1,4)\}$,

3 - $\{(2,0)\to(1,1)\,\&\,(2,0)\to(1,5)\}$.

The behavior of the scattering rate for the different types of transitions with a change in the transferred energy can be explained as follows.

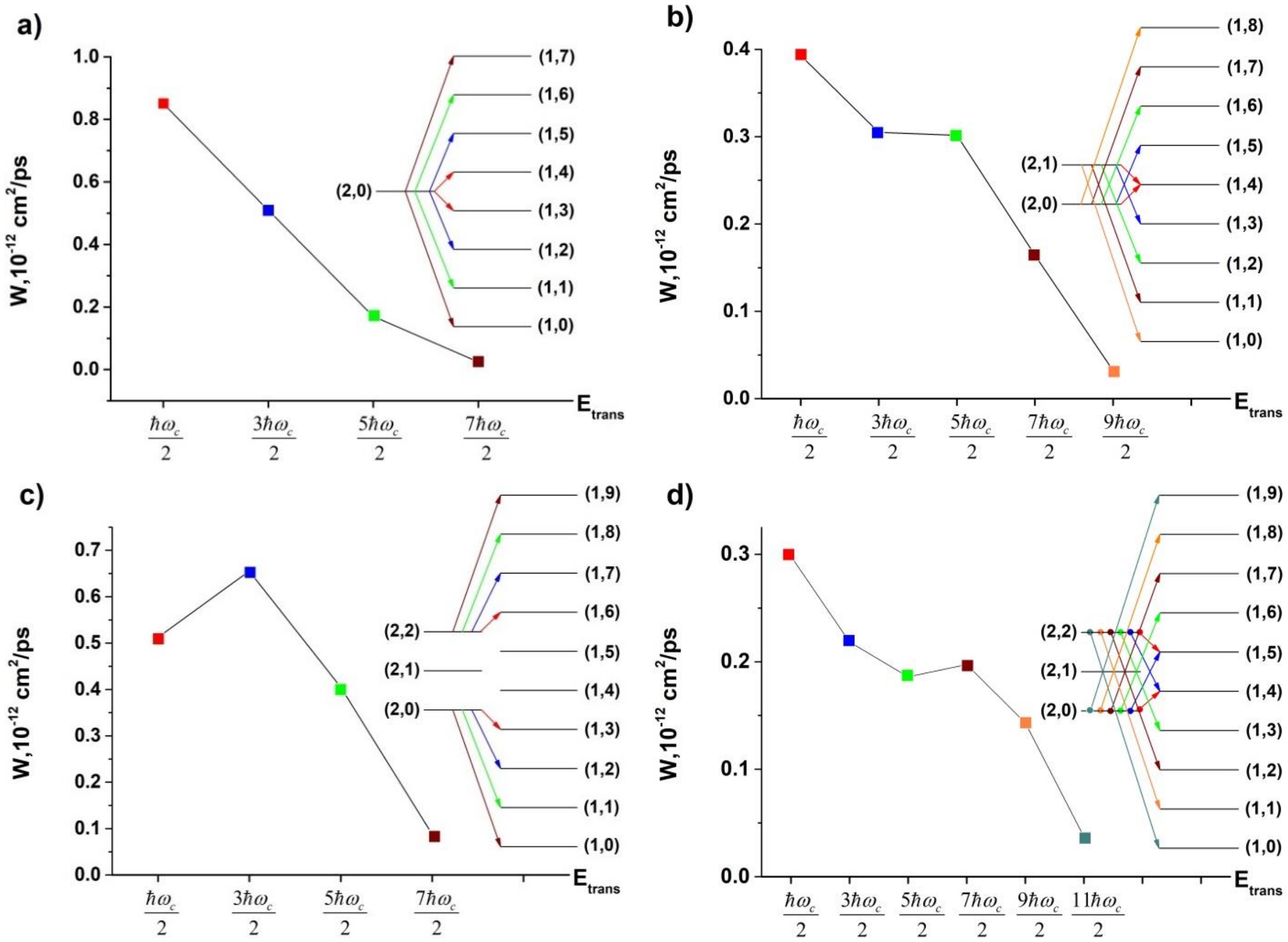

Figure 9. Dependence of the rate of type I interdimensional transitions on the transferred energy. The magnetic field B = 3.5 T corresponds to the situation where the Landau level (2,0) lies between the Landau levels (1,3) and (1,4).

In resonance, the transition rate is determined not directly by the change in electron energy, but rather by the difference in the wave function (2) of each electron before and after scattering. Since the structure potential and, consequently, the subband energy levels $\varepsilon_\nu$ and wave functions $\varphi_\nu(z)$ are fixed, the difference in the wave functions of the initial and final states of a transition

$$\left\{(i_i,n_i)\rightarrow\left(\nu_f,n_f\right)\&\left(\nu_j,n_j\right)\rightarrow\left(\nu_g,n_g\right)\right\}$$

of this type is determined by the difference in the numbers of the Landau levels

$$\Delta n_1=\left|n_f-n_i\right| \text{ and } \Delta n_2=\left|n_g-n_j\right|,$$

in the subbands to which these levels belong. The smaller this difference is, the closer the wave functions of the initial and final states are. In the case of $\Delta n_\alpha = 0$ the difference of the wave functions of the initial and final states of the $\alpha$-th electron of the pair is minimal, these wave functions differ only by the subband component.

In the case of intrasubband transitions of type I $\Delta n_1 = \Delta n_2 = \Delta n$, and the transferred energy $E_{trans} = \hbar\omega_c \cdot \Delta n$. Therefore, the increase in $E_{trans}$, that is $\Delta n$, the difference between the wave functions of the initial and final states of each electron of the interacting pair increases, and, as a consequence, the transition rate decreases. This explains the observed monotonic dependence of the intrasubband transition rates on $E_{trans}$. The aforementioned reason also clarifies why the difference between the rates decreases with increasing $\Delta n$. This occurs because the more zeros the wave function has, the smaller the difference becomes when an additional zero is added.

Therefore, as $E_{trans}$ as well as $\Delta n$ enhances, the difference between the wave functions of the initial and final states of each electron in the interacting pair increases, and, as a consequence, the transition rate decreases. This explains the observed monotonic dependence of the intrasubband transition rates on $E_{trans}$. This reason also explains why the difference between the rates decreases with increasing $\Delta n$. This is because the more zeros the wave function has, the smaller this difference is when another zero is added.

In the case of type I intersubband transitions $\Delta n_1 \neq \Delta n_2$. This is clearly evident in the diagrams of transitions shown in Figure 9 as insets. For example, for the transition

$$\{(2,0) \to (1,3) \,\&\, (2,0) \to (1,4)\}$$

the Landau level number change for one electron is $\Delta n_1 = |3-0| = 3$, and for the other is $\Delta n_2 = |4-0| = 4$. For intersubband transitions of type I, we have introduced a number so that $E_{trans} = \hbar\omega_c \cdot \delta n$. The value $\delta n$ gives us the number of Landau levels through which the electron "jumps" at scattering (see the schematics in Figs. 6 and 7). However, since the subbands are shifted relative to each other, $\Delta n_1$ and $\Delta n_2$ do not coincide with this number of levels.

The relationship of $\Delta n_1$ and $\Delta n_2$ to $\delta n$ is described by expressions (31) and (32) in the case of strict coincidence of the upper and lower subband levels. For resonance, when the levels of the upper subband are positioned midway between the Landau levels of the lower subband, expressions (35) and (36) apply. From these expressions, as well as the transition diagrams shown in the insets of Figures 9, it is evident that as the transferred energy $E_{trans}$ and $\delta n$ increases, the difference in the Landau level numbers for one electron also increases, which contributes to a decrease in the transition rate. Meanwhile, this difference initially decreases for the second electron in the scattering pair, increasing the transition rate. Once the value reaches zero at $\delta n = p$ or $\delta n - \frac{1}{2} = p$, it starts to increase, acting together with the first electron to lower the matrix element. This increase occurs in conjunction with the first electron, and both influence the matrix element in a direction that reduces its value. As a result, we observe that if $\delta n \leq p$ the electrons in the scattering pair behave oppositely regarding their impact on the transition rate, as the value $\delta n$ increases. This leads to a complex relationship between $E_{trans}$ and the intersubband transition rate of type I.

In the case of type III intersubband transitions, the dependence on $E_{trans}$ behaves similarly, though it may be more pronounced. For instance, in type III transitions, it is possible for both $\Delta n_1$ and $\Delta n_2$ to decrease simultaneously as the transferred energy $E_{trans}$ increases, while the transition rate significantly and consistently rises with increasing $E_{trans}$. This scenario is illustrated in Figure 10a, where it is evident that the scattering rate increases nearly tenfold as energy increases by almost an order of magnitude with increasing $E_{trans}$ .

However, this dependence is not a universal trend. Here, as in the first case, the nature of the dependence is influenced in complex ways by the initial states of the electrons in the scattering pair. In some instances, the transition rate may decrease monotonically and somewhat slowly (see Fig. 10b). The situation may differ for other initial states of the electrons, where the rate can initially increase significantly before rapidly decreasing (see Fig. 10c).

The reason for this behavior here is the same as for the intersubband transitions of type I. The magnitude of the transition rate is determined by $\Delta n_1$ and $\Delta n_2$. Since the subbands are shifted relative to each other, there is no unambiguous relationship between $\Delta n_1$ , $\Delta n_2$ and the transferred energy $E_{trans}$. The dependence of $E_{trans}$ on $\Delta n_1$ and $\Delta n_2$ is specific for each of the different initial states. At the same time $\Delta n_1$ and $\Delta n_2$ may not coincide and can change up or down with increasing $E_{trans}$.

We arrive at an important conclusion: it is impossible to assess the relative magnitude of intersubband transition rate based solely on the transferred energy $E_{trans}$. Additionally, since both $\Delta n_1$ and $\Delta n_2$ differ, and can change variably depending on the initial states of the electrons, there is no explicit relationship between the transition rate and $\Delta n_1$ and $\Delta n_2$, which holds for all transitions of this type. As a result, it is not feasible to determine in advance which intersubband electron-electron scattering processes are the most significant and should be considered in kinetic assessments, and which ones are less intense and can therefore be ignored. Thus, for an accurate description of the kinetics, it is essential to compute the "complete" matrix of intersubband transition rates.

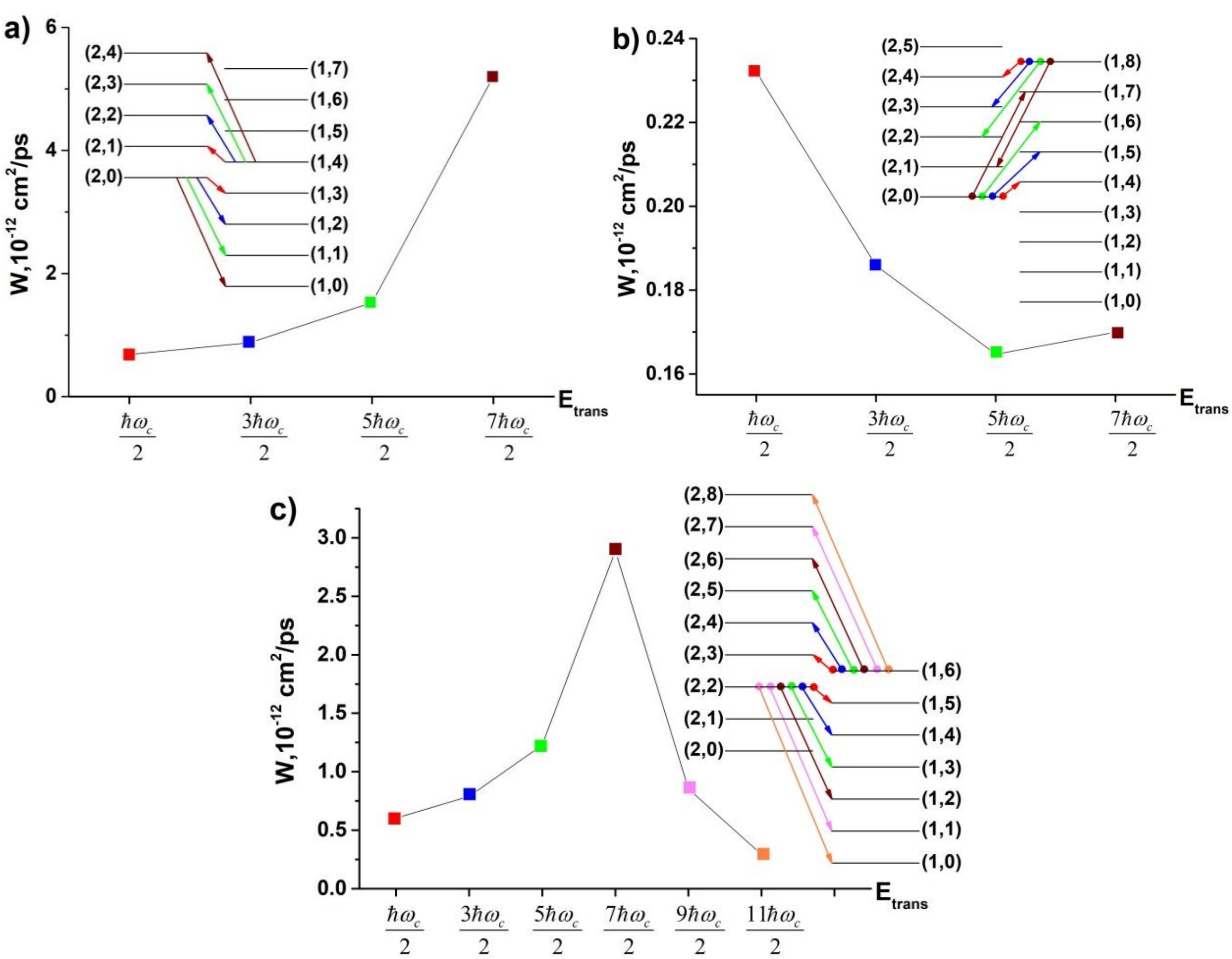

Figure 10. Dependencies of the type III transition rate on transferred energy $E_{trans}$. Magnetic field B=3.5 T.

This is illustrated in Fig. 11, which shows the dependence of the excitation energy of the electronic subsystem on time during its selective excitation to the Landau level (2,0), which is situated below the optical phonon energy level. Curve 1 is calculated by considering all electron-electron scattering processes occurring among electrons at Landau levels below the energy of the optical phonon. In the calculation of Curve 2, all intersubband electron-electron scattering processes are disregarded, except for transitions $\{(2,0)\rightarrow(1,3)\,\&\,(2,0)\rightarrow(1,4)\}$ to the nearest Landau levels of electrons that are initially at the same Landau level (2,0). As demonstrated, including the complete set of intersubband electron-electron scattering rates results in a significant acceleration of energy relaxation, reducing the relaxation time by more than 50%.

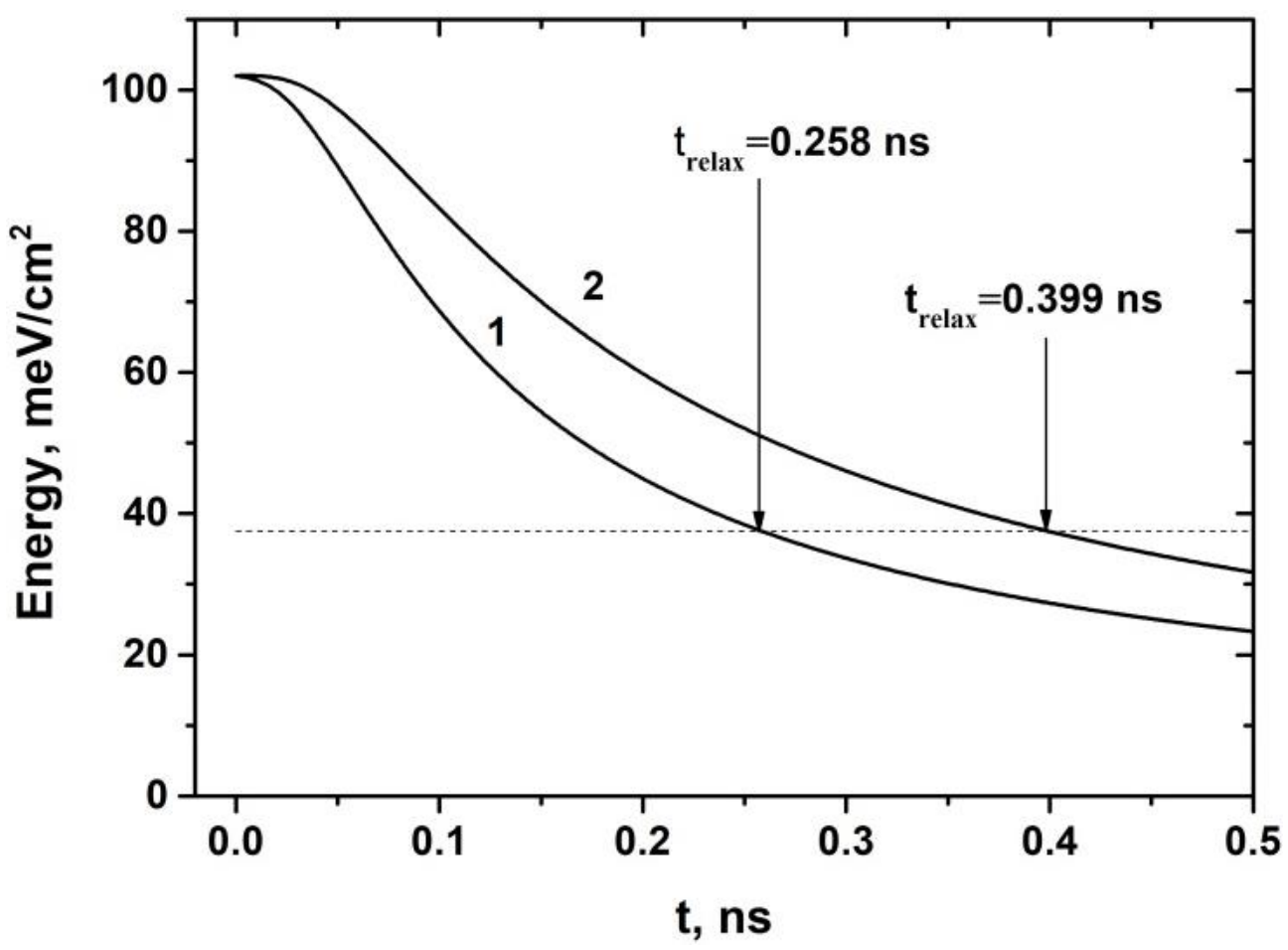

Figure 11. Dependence of the excitation energy of the electron subsystem on time calculated with "full" matrix of electron-electron scattering rates (curve 1) and neglecting all intersubband transitions except for transitions $\{(2,0)\rightarrow(1,3)\,\&\,(2,0)\rightarrow(1,4)\}$ (curve 2). At the initial moment of time t=0, only two Landau levels are populated - level (2,0) with a concentration of $5\cdot10^9$ $cm^{-2}$ and level (1,0) with a concentration of $10^{10}$ $cm^{-2}$. The magnetic field V=3.5 Tesla

In the case of transitions of type I, the interacting electrons are in the same subband. Accordingly, $\Delta E_{init}=\hbar\omega_c\left|\Delta N\right|$, and the rates for transitions of this type slowly decrease with $\Delta E_{init}$ (Fig. 12).

In the cases of type II intrasubband transitions and type III intersubband transitions, the electrons start in different subbands that are shifted relative to each other. As a result, there is no unambiguous dependence of $\Delta E_{init}$ on $\left|\Delta N\right|$, which also causes the transition rate to lack a simple, single-valued relationship with $\Delta E_{init}$.

To illustrate the above, let us consider intrasubband transitions of type II at the value of the magnetic field when the Landau level (2,0) coincides with the level (1,3). Let us consider the transition $\{(1,1+\Delta N)\to(1,\Delta N)\,\&\,(2,0)\to(2,1)\}$. From the scheme of transitions in the inset to Figure 13, we can see that the magnitude of scattering rate decreases when $\Delta N$ is increased from 1 to 3, and increases when $\Delta N$ is increasing further. As a result, for the first part of the transition, their rate decreases with increasing $\Delta N$, and for the second part, it increases.

It should be noted that the dependence on $\Delta N$ is weak for all types of transitions.

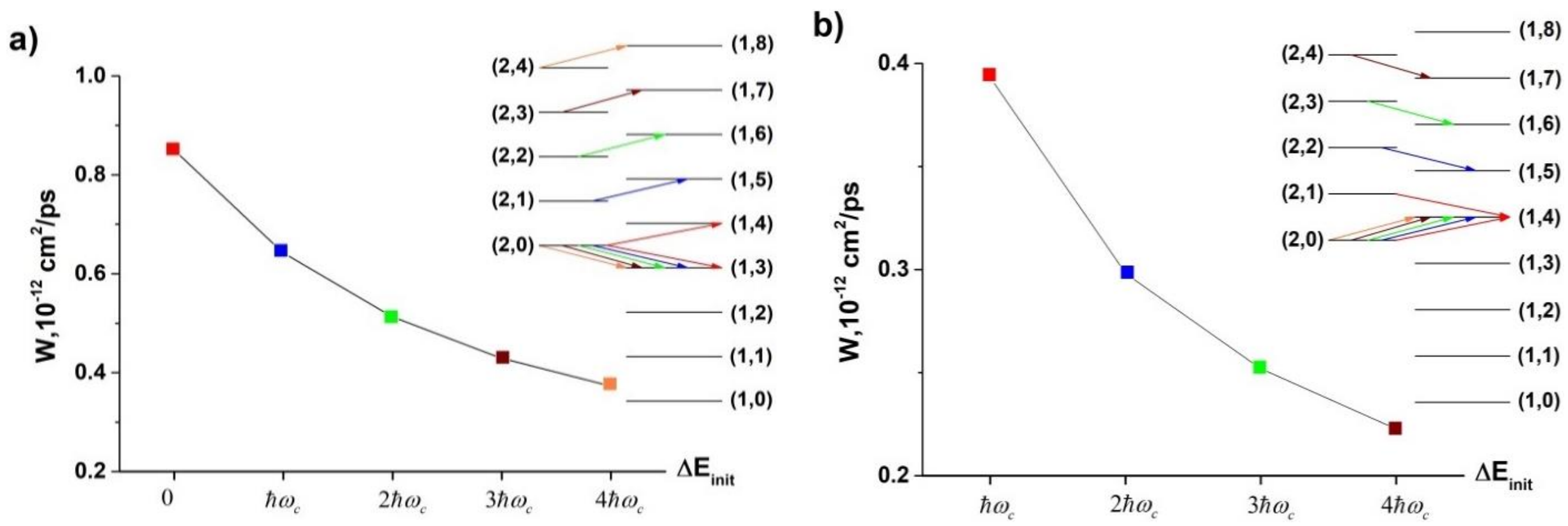

Figure 12. Dependence of the rate of intersubband scattering of type I on the difference of initial values of the energy of electrons of the interacting pair $\Delta E_{init}$. Magnetic field $B=3.5$ T.

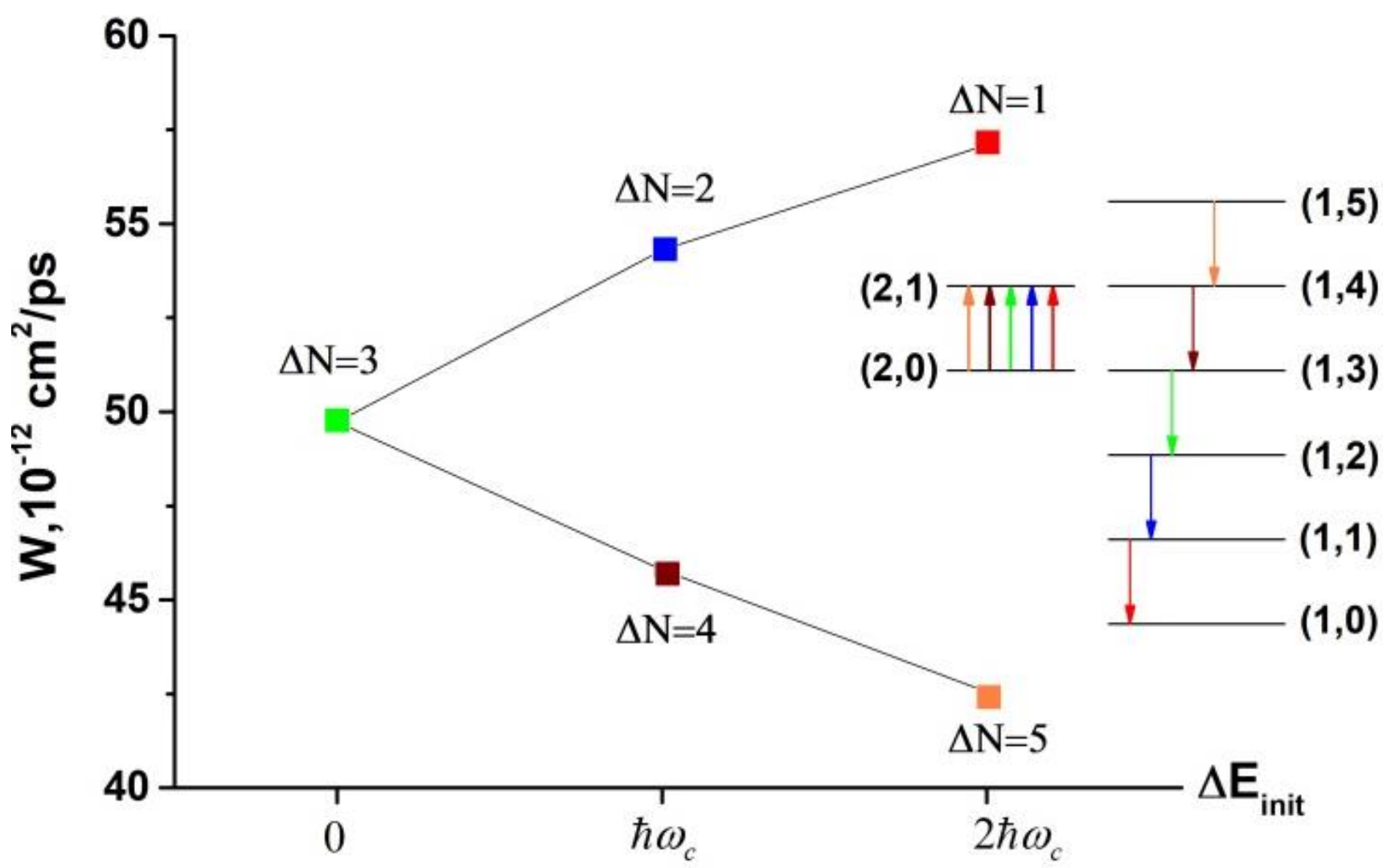

Figure 13. Dependence of the type II intrasubband scattering rate on the difference in the electron energy of the interacting pair before scattering. The magnetic field $B=4.1$T corresponds to the resonance when the Landau level (2,0) coincides with the Landau level (1,3).

For type III intersubband transitions, the resonance conditions have the same form (22) as for intrasubband transitions. Therefore, the dependence of the rate of these transitions on the magnetic field strength is smooth (Fig. 14). The rate of type III transitions can reach values several times greater than the rate of type I intersubband transitions, while remaining an order of magnitude lower than the rate of intrasubband transitions.

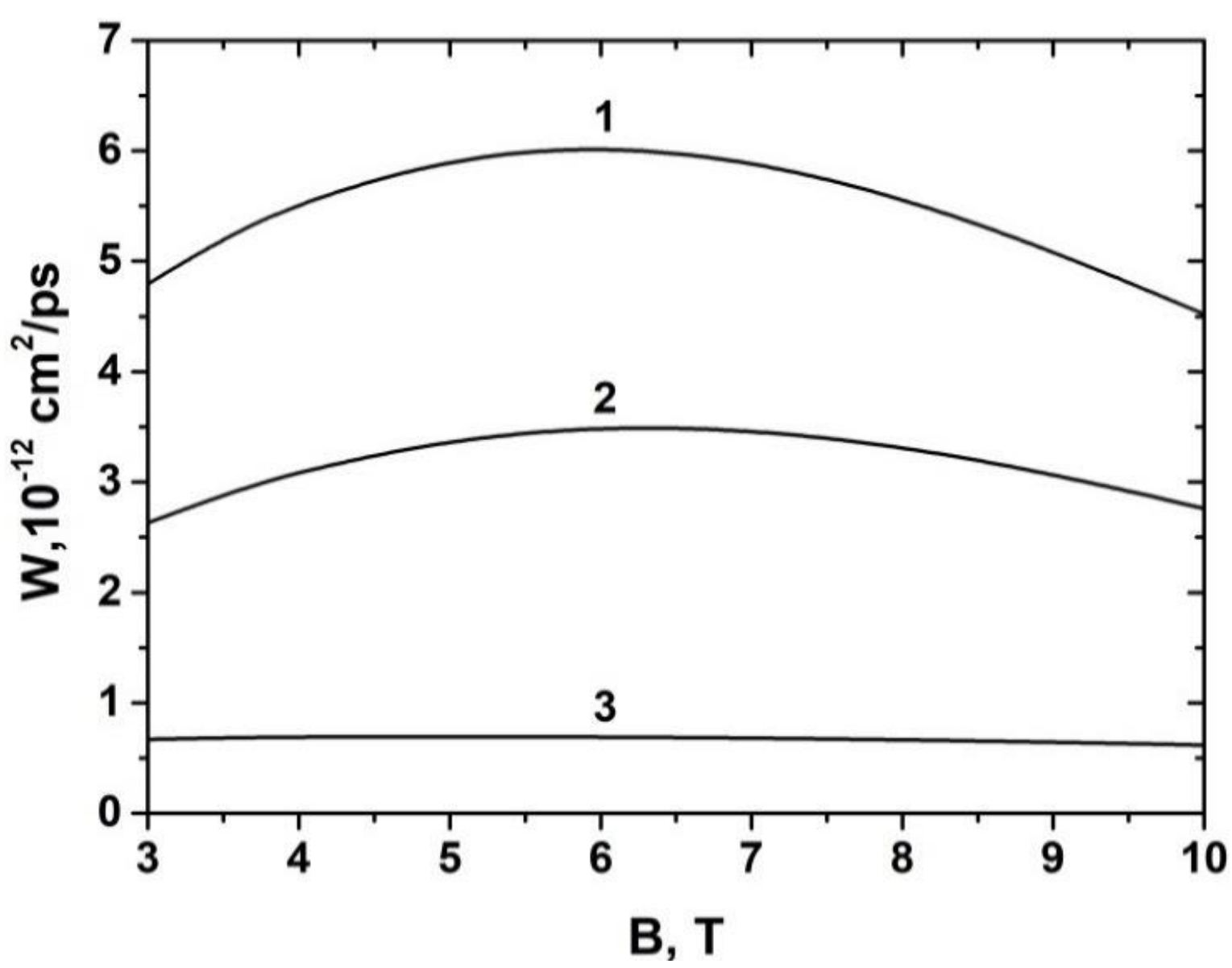

Figure 14. Dependence of the rate of type III transitions on the magnetic field strength.
1- $\{(2,0)\rightarrow(1,0)\,\&\,(1,4)\rightarrow(2,4)\}$, 2- $\{(2,2)\rightarrow(1,2)\,\&\,(1,6)\rightarrow(2,6)\}$,
3-$\{(2,0)\rightarrow(1,3)\,\&\,(1,4)\rightarrow(2,1)\}$.

In type II intersubband transitions, resonance occurs when the Landau levels of the upper subband align with those of the lower subband. This alignment happens at specific magnetic field values defined by expression (29). As a result, the dependence of these transition rates on the magnetic field is a resonance peak.

The obtained expressions for the scattering rate enabled us to establish a selection rule for type II intersubband transitions. This rule is defined by the symmetry of the quantum well's potential profile. In structures that have a symmetric profile, the scattering rate is zero for all transitions

$$\left\{(\nu_i,n_i)\rightarrow(\nu_f,n_f)\,\&\,(\nu_j,n_j)\rightarrow(\nu_g,n_g)\right\}$$

with an odd sum of subband numbers $\nu_i+\nu_j+\nu_g+\nu_f$.

Let us examine the change in the function $R_{\nu_i,\nu_j,\nu_g,\nu_f}(z_1)$ defined by expression (12) when the sign of the argument is altered.

$$R_{\nu_i,\nu_j,\nu_g,\nu_f}\left(-z_1\right)=\int_{-\infty}^{+\infty} dz_2\varphi_{\nu_i}(z_2)\varphi_{\nu_j}(z_2+2z_1)\overset{*}{\varphi}_{\nu_g}(z_2+2z_1)\overset{*}{\varphi}_{\nu_f}(z_2). \tag{37}$$

Substituting the variable in the integral $z_2 \to -z_2$, we obtain

$$R_{\nu_i,\nu_j,\nu_g,\nu_f}\left(-z_1\right)=\int_{-\infty}^{+\infty} dz_2\varphi_{\nu_i}(-z_2)\varphi_{\nu_j}(-z_2+2z_1)\overset{*}{\varphi}_{\nu_g}(-z_2+2z_1)\overset{*}{\varphi}_{\nu_f}(-z_2). \tag{38}$$

In structures with symmetric potential profile, the subband wave functions are divided by parity

$$\varphi_\nu(-z)=\left(-1\right)^{\nu+1}\varphi_\nu(z). \tag{39}$$

Substituting (39) into (38), we obtain

$$R_{\nu_i,\nu_j,\nu_g,\nu_f}\left(-z_1\right)=\left(-1\right)^{\nu_i+\nu_j+\nu_g+\nu_f}R_{\nu_i,\nu_j,\nu_g,\nu_f}\left(z_1\right). \tag{40}$$

When the sum $\nu_i+\nu_j+\nu_g+\nu_f$ is odd, the function $R_{\nu_i,\nu_j,\nu_g,\nu_f}(z_1)$ itself is also odd. As a result, the integrand in integral (28), which defines the function $G_{\nu_i,\nu_j,\nu_g,\nu_f}(\gamma;y)$, is likewise odd. Therefore, the function, and consequently the rates of all transitions where the sum of subband numbers $\nu_i+\nu_j+\nu_g+\nu_f$ is odd, are equal to zero.

Thus, in the case when the sum $\nu_i+\nu_j+\nu_g+\nu_f$ is odd, the function $R_{\nu_i,\nu_j,\nu_g,\nu_f}(z_1)$ is also odd, and, as a consequence, the integrand in the integral (28) defining the function $G_{\nu_i,\nu_j,\nu_g,\nu_f}(\gamma;y)$ is odd. Therefore, the function $G_{\nu_i,\nu_j,\nu_g,\nu_f}(\gamma;y)$ and, consequently, the rates of all transitions in which the sum of subband numbers $\nu_i+\nu_j+\nu_g+\nu_f$ is odd are equal to zero.

For type II transitions, when one of the electrons moves to a neighboring subband $\nu_i=\nu_j=\nu_f=\nu_g\pm1$, and $\nu_i+\nu_j+\nu_g+\nu_f=4\nu_i\mp1$ is correspondingly odd.

Therefore, type II transitions between neighboring subbands are forbidden in structures with a symmetric profile. However, in the case of transitions between subbands located further apart (for example, between the first and third), the rate of these transitions is non-zero (Fig. 15). Nonetheless, this intensity is nearly two orders of magnitude lower than that of type I intersubband transitions. This difference is also attributed to the structure of wave functions in quantum wells with a symmetric potential profile, as illustrated in Figure 16, which displays the functions $R_{\nu_i,\nu_j,\nu_g,\nu_f}(z_1)$ associated with different types of transitions.

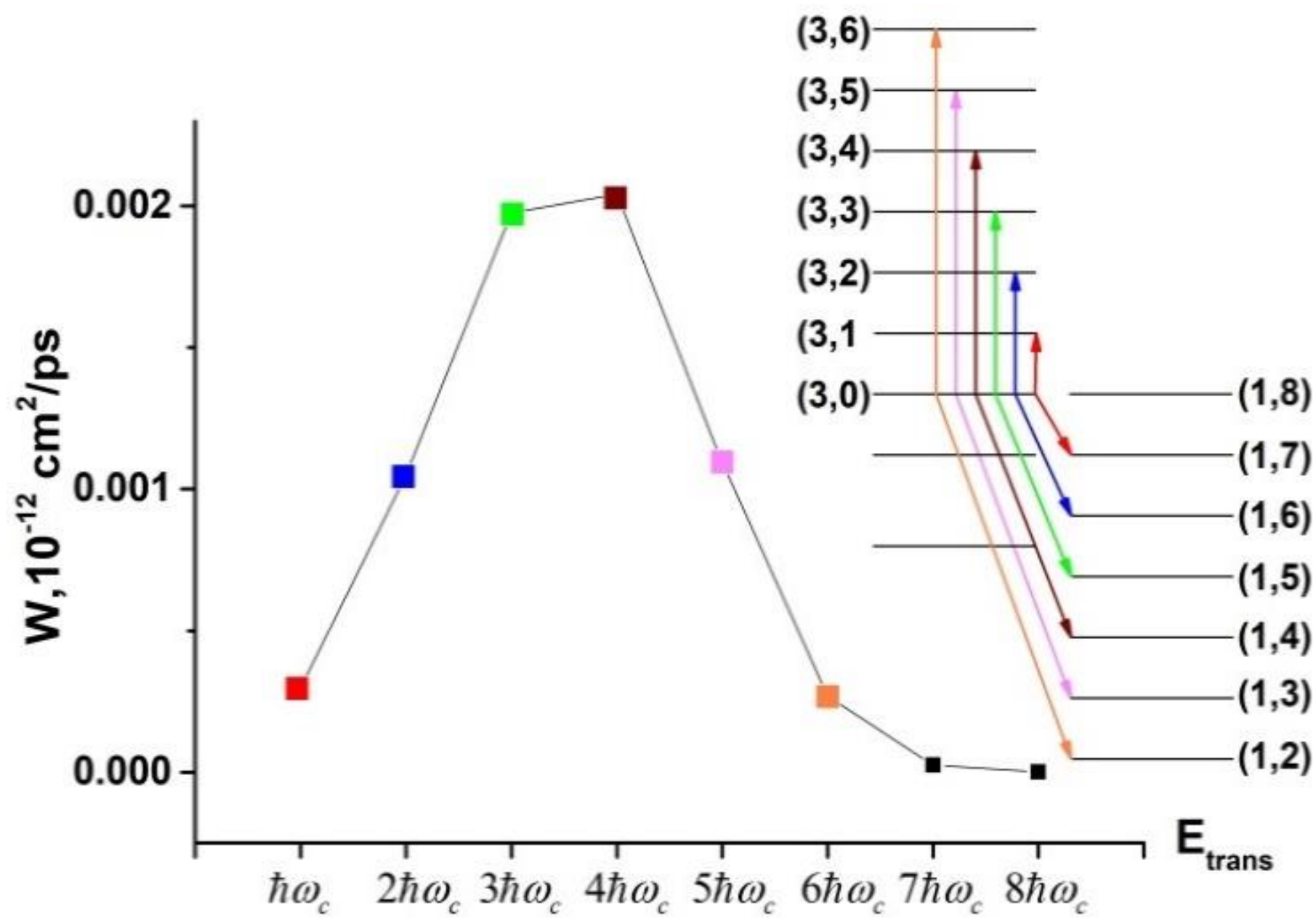

Figure 15. Dependence of the type II intersubband transition rate on the energy transferred to one electron in the scattering event. The magnetic field B=4.2 T corresponds to the resonance when the Landau levels (3,0) and (1,8) coincide.

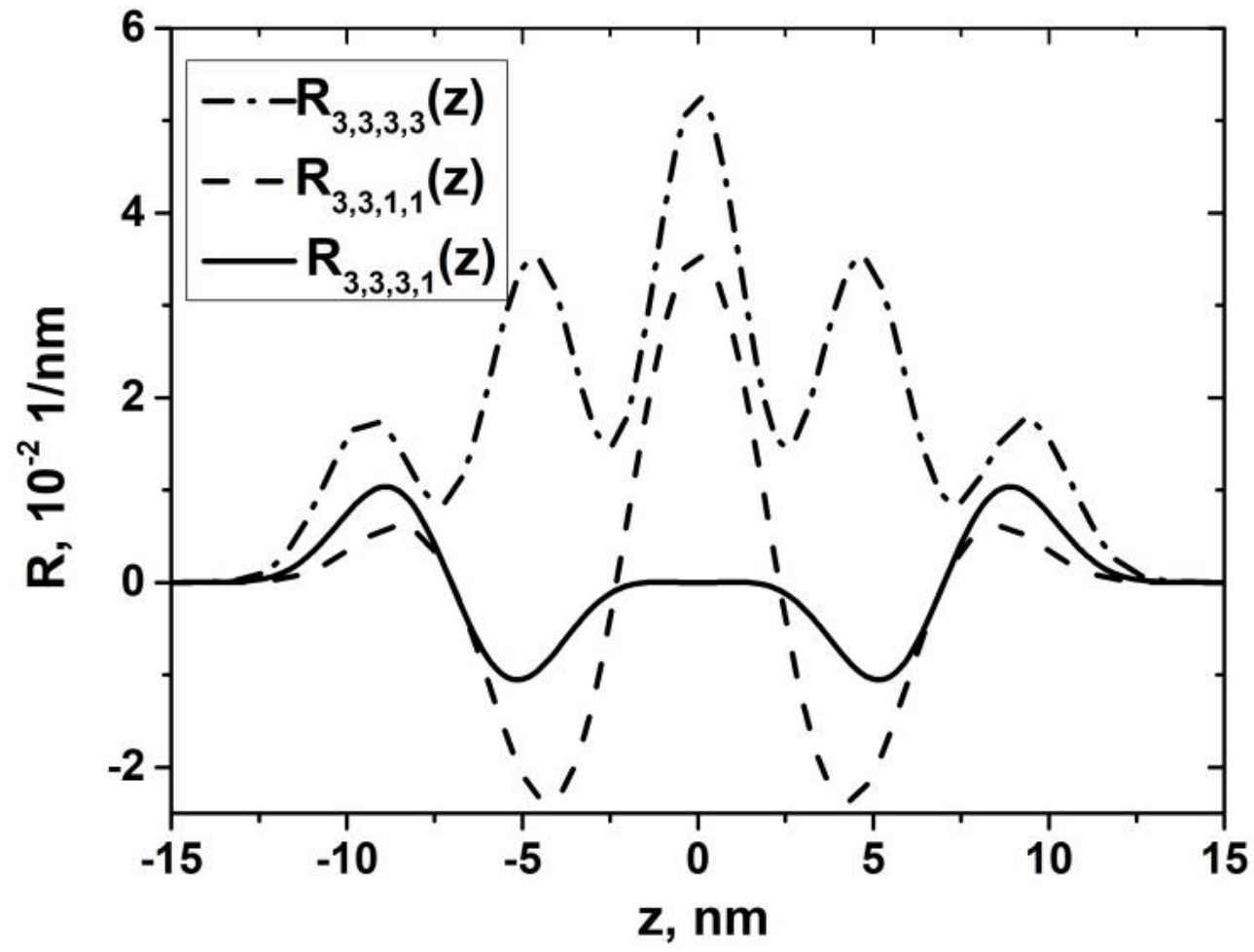

Figure 16. Function $R_{\nu_i,\nu_j,\nu_g,\nu_f}(z)$ for transitions of different types - intrasubband transitions in the third subband (dashed-dotted curve), intersubband transitions of type I (dashed curve) and type II (solid curve) from the third to the first subband.

In structures with asymmetric potential profiles, the subband wave functions cease to have a certain parity ($\varphi_\nu(-z) \neq \varphi_\nu(z)$), resulting in non-zero probabilities for intersubband transitions of type II. Consequently, the magnitude, behavior, and relative contribution of type II transitions depend significantly on the symmetry of the structure potential profile. Therefore, it is crucial to consider these transitions for each specific type of structure.

## 6. Effect of magnetic field orientation.

In the case when the magnetic field has a component $B_\parallel$ parallel to the layers of the quantum well, expressions (1) give the following resonance condition for the transition

$$\Delta\varepsilon_{if} + \delta\varepsilon_{if}\left(B_\parallel\right) + \Delta\varepsilon_{jg} + \delta\varepsilon_{jg}\left(B_\parallel\right) + \left(n_i - n_f + n_j - n_g\right)\cdot\hbar\omega_\perp = 0, \quad (41)$$

where

$$\delta\varepsilon_{if}\left(B_\parallel\right) = \frac{e^2}{2mc^2}\left[\left(\delta z\right)^2_{\nu_i} - \left(\delta z\right)^2_{\nu_f}\right]\cdot B^2_\parallel \quad (42)$$

is the change in the distance between the $\nu_i$-th and $\nu_f$-th subbands due to the shift of each of the subbands as a whole caused by $B_\parallel$. According to the oscillator theorem, the wave function of the ν-th energy level contains (ν+1) zeros. Therefore, as ν increases, the complexity of the wave function also increases, leading to a larger root mean square (RMS) fluctuation $\left(\delta z\right)_\nu$ in the z-coordinate. Consequently, $\delta\varepsilon_{if}\left(B_\parallel\right) > 0$ if $\nu_i > \nu_f$, so when the magnetic field component $B_\parallel$ increases, it causes a greater distance between the upper and lower subbands, as illustrated in Figure 17.

The shift of the subbands does not change their structure, meaning it does not affect the resonance condition as long as each electron remains in its subband during the scattering event. This applies to type I and type II intrasubband transitions. For these transitions $\Delta\varepsilon_{if} = \Delta\varepsilon_{jg} = 0$ and $\delta\varepsilon_{if} = \delta\varepsilon_{jg} = 0$. As a result, the resonance condition in a tilted magnetic field remains the same as in a magnetic field that is perpendicular to the layers of the structure.

Neither does the resonance condition change for intersubband transitions of type III. Although in this case $\delta\varepsilon_{if}\left(B_\parallel\right)$ is nonzero for each of the electrons, however, $\Delta\varepsilon_{if} = -\Delta\varepsilon_{jg}$ and $\delta\varepsilon_{if}\left(B_\parallel\right) = -\delta\varepsilon_{jg}\left(B_\parallel\right)$.

In the case of intersubband transitions of type I and type II, component $B_{\|}$ leads to a shift of their resonances (Fig. 18). For the intersubband transition

$$\left\{\left(\nu_i,n_i\right)\to\left(\nu_f,n_f\right)\&\left(\nu_i,n_j\right)\to\left(\nu_f,n_g\right)\right\}$$

of type I, the condition (41) takes the following form

$$2\Delta\varepsilon_{if}+2\delta\varepsilon_{if}\left(B_{\|}\right)-\frac{2\Delta\varepsilon_{if}}{\hbar\omega_{\perp}^{(0)}}\cdot\hbar\omega_{\perp}=0, \qquad (43)$$

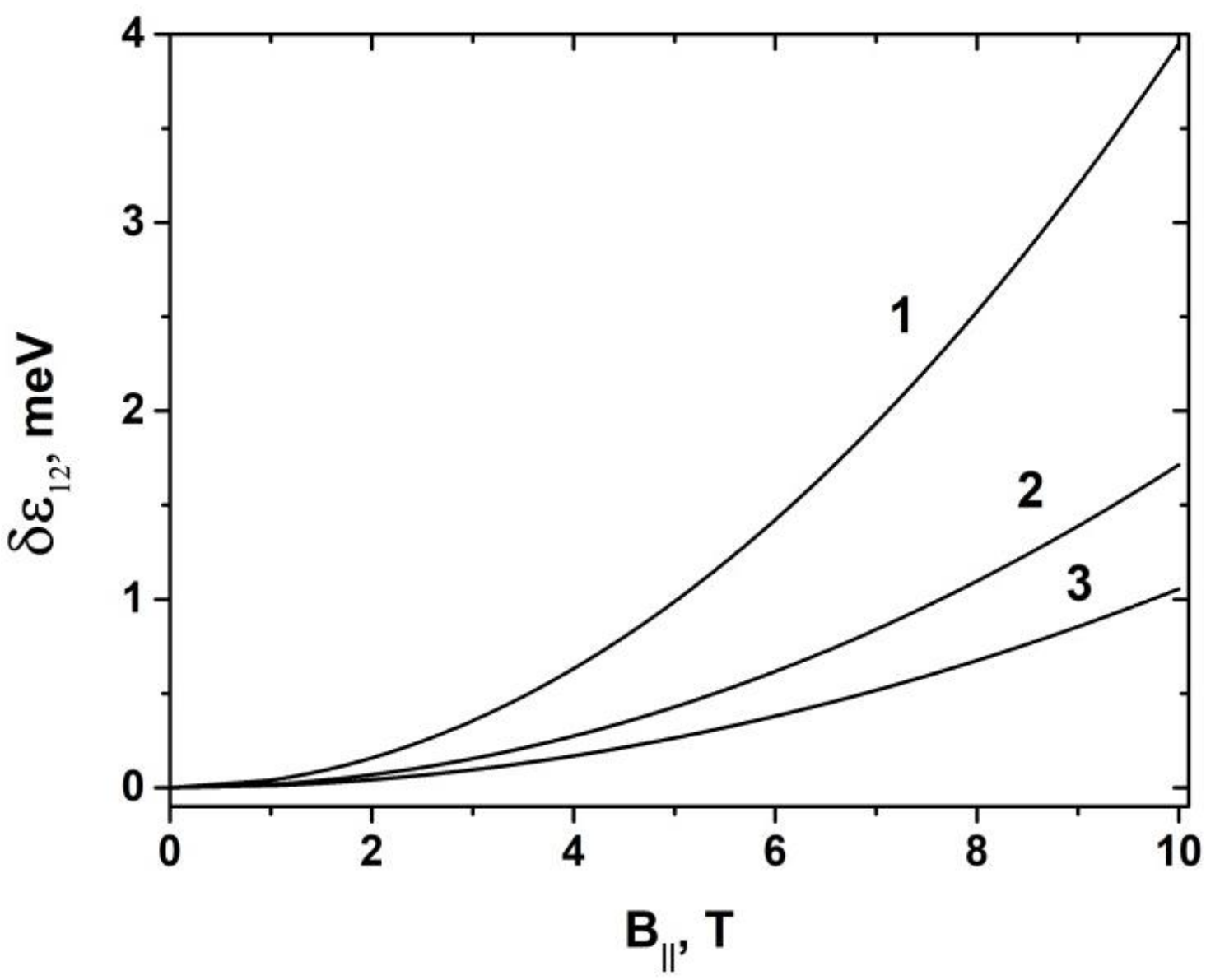

Figure 17. Dependence of the distance between the two lowest subbands in a quantum well on the magnetic field component parallel to the layers. Different curves correspond to different values of the quantum well width *a*: 1 - *a*=25 nm; 2 - *a*=15 nm; 3 - *a*=10 nm. The data are given for the $GaAs/Al_{0.3}Ga_{0.7}As$ quantum well.

Here, we use the resonance condition (25) for this transition in perpendicular magnetic field, $\omega_{\perp}^{(0)}=\dfrac{eB_{\perp}^{(0)}}{mc}$, $B_{\perp}^{(0)}$ - the resonant value of perpendicular magnetic field when $B_{\|}=0$. From (43), we find the resonant value of the magnetic field component $B_{\perp}$, at which the resonance takes place, if $B_{\|}\neq 0$:

$$\frac{B_{\perp}}{B_{\perp}^{(0)}}=1+\frac{\delta\varepsilon_{if}\left(B_{\|}\right)}{\Delta\varepsilon_{if}} \qquad (44)$$

Accordingly, the relative shift

$$\frac{\delta B_{\perp}}{B_{\perp}^{(0)}} = \frac{B_{\perp} - B_{\perp}^{(0)}}{B_{\perp}^{(0)}} = \frac{\delta\varepsilon_{if}\left(B_{\parallel}\right)}{\Delta\varepsilon_{if}} . \tag{45}$$

The relative shift in the magnetic field value at which the transition resonance occurs is independent of the Landau level numbers. It is equal to the relative change in the intersubband distance. As illustrated in Figure 19, the relative shift increases significantly as the width of the quantum well increases. This occurs because $\Delta\varepsilon_{if}$ decreases while $\delta\varepsilon_{if}\left(B_{\parallel}\right)$ increases with an increase in the quantum well width. The increase in $\delta\varepsilon_{if}\left(B_{\parallel}\right)$ is caused by the growth of the localization area of the wave function, which, in turn, leads to a larger standard deviation $\delta z$ for each of the subbands.

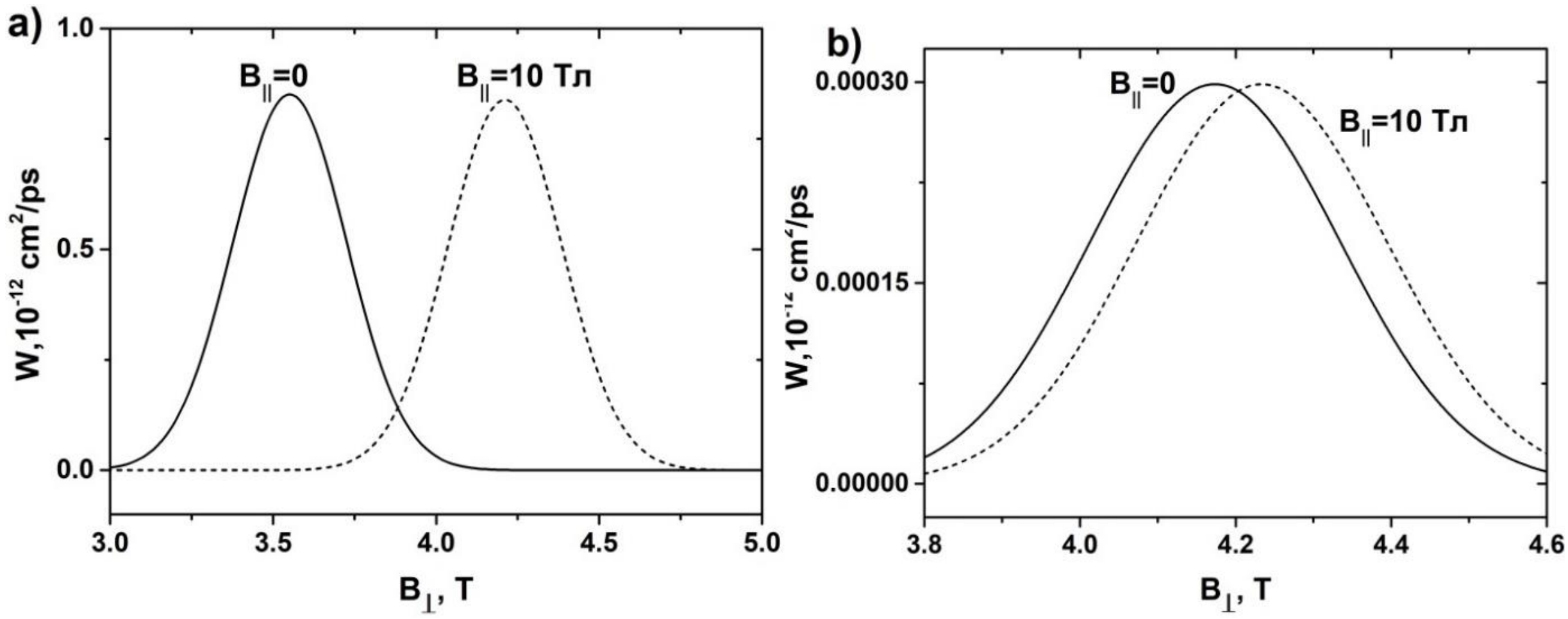

Figure 18. Effect of the magnetic field component $B_{\parallel}$ parallel to the quantum well layers on the dependence of the electron-electron scattering rate for intersubband transitions of type I (a) and type II (b) on the magnetic field component $B_{\perp}$ perpendicular to the quantum well layers. The solid curve - $B_{\parallel} = 0$, the dashed curve - $B_{\parallel} = 10$ T. The data are given for the transition $\left\{(2,0) \to (1,3) \,\&\, (2,0) \to (1,4)\right\}$ (a) and $\left\{(3,0) \to (3,1) \,\&\, (3,0) \to (1,7)\right\}$ (b) in a GaAs/$Al_{0.3}Ga_{0.7}As$ quantum well of 25 nm width.

To estimate the effect, let us consider deep levels in a rectangular quantum well with width *a*, where most of the wave function is confined within the well,

$$\hbar / \sqrt{2m_w(U_0 - \varepsilon)} \ll a .$$

Thus, we can neglect the penetration of the wave function into the barrier and consider the quantum well as infinitely deep. Then we obtain

$$\left(\delta z\right)_{\nu}^{2}=\frac{a^{2}}{12}\left(1-\frac{6}{\left(\pi\nu\right)^{2}}\right), \tag{46}$$

and

$$\delta\varepsilon_{if}\left(B_{\parallel}\right)=\frac{e^{2}}{4\pi^{2}mc^{2}}\left\lfloor\frac{1}{\nu_{f}^{2}}-\frac{1}{\nu_{i}^{2}}\right\rfloor\cdot a^{2}\cdot B_{\parallel}^{2} \tag{47}$$

Correspondingly, for the relative magnitude of the resonance shift, we have

$$\frac{\delta B_{\perp}}{B_{\perp}^{(0)}}=\frac{B_{\perp}-B_{\perp}^{(0)}}{B_{\perp}^{(0)}}=\frac{\delta\varepsilon_{if}\left(B_{\parallel}\right)}{\Delta\varepsilon_{if}}\frac{e^{2}}{2\pi^{4}c^{2}\hbar^{2}}\frac{1}{\nu_{i}^{2}\nu_{f}^{2}}\cdot a^{4}\cdot B_{\parallel}^{2}. \tag{48}$$

Thus, the relative shift of the resonance grows as the fourth power of the quantum well width.

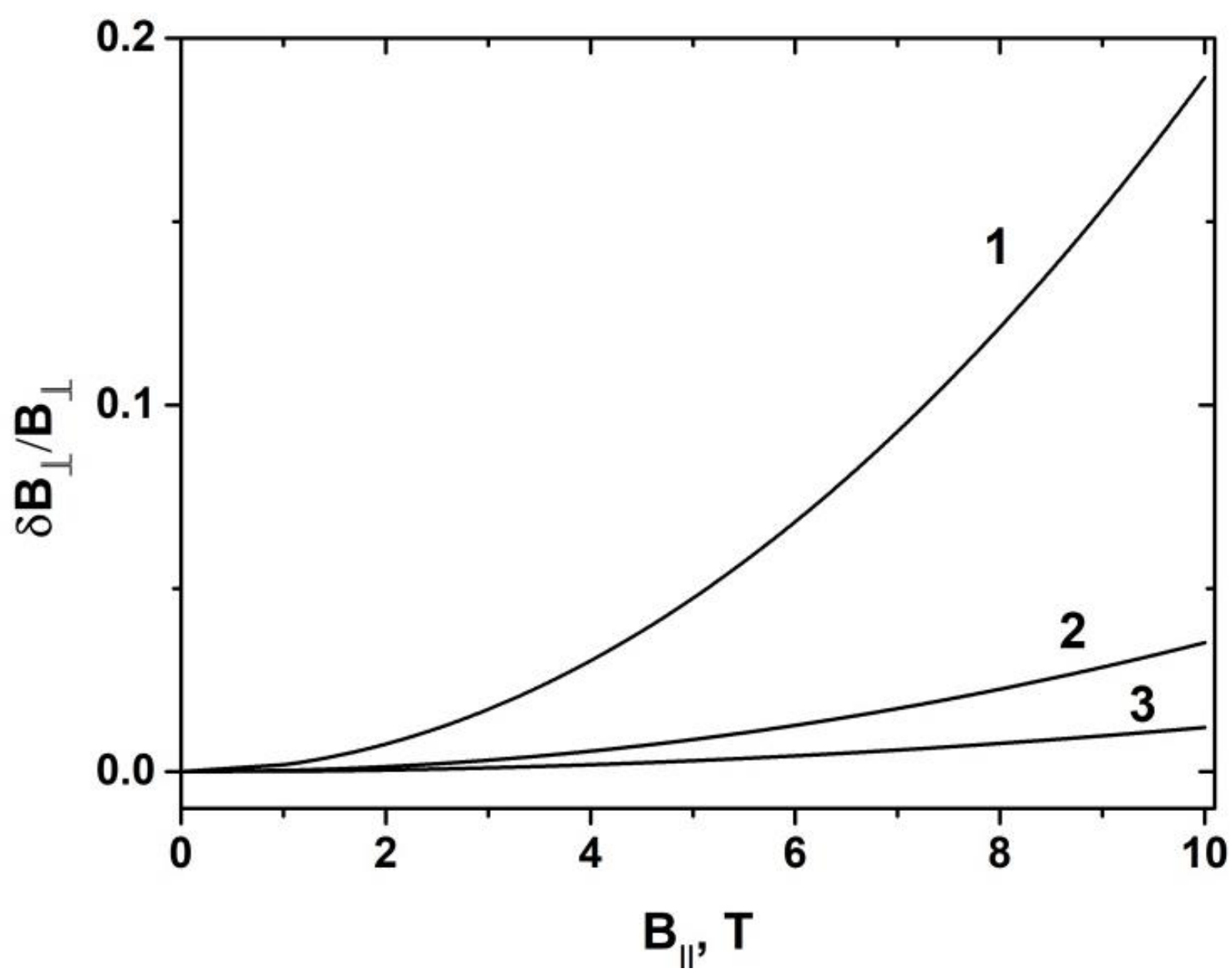

Figure 19. Dependence of the relative shift of the type I transition resonance between the lowest subbands on the magnetic field component $B_{\parallel}$ parallel to the layers of the quantum well. Different curves correspond to different values of the width *a* of the quantum well: 1 - *a*=25 nm; 2 - *a*=15 nm; 3 - *a*=10 nm.

It follows from (45) that the resonance at the transition from the upper subband to the lower one shifts in the direction of higher magnetic fields and has the value of

$$\delta B_{\perp}=\frac{mc}{e\hbar}\frac{2\delta\varepsilon_{if}\left(B_{\parallel}\right)}{n_{f}+n_{g}-n_{i}-n_{j}}=\frac{e}{\hbar c}\frac{\left\lfloor\left(\delta z\right)_{\nu_{i}}^{2}-\left(\delta z\right)_{\nu_{f}}^{2}\right\rfloor}{n_{f}+n_{g}-n_{i}-n_{j}}\cdot B_{\parallel}^{2} \tag{49}$$

As can be seen from (49), the magnitude of the shift is proportional to the increment in the intersubband distance and, accordingly, grows with increasing quantum well width approximately as the square of the width (Fig. 20).

The shifts of resonances depend rather strongly (quadratically) on $B_{\|}$, which makes it possible for this component to provide a significant effect on both the magnitude of type I intersubband scattering rates and the ratio of rates of various transitions. As a result, some transitions are considerably suppressed, while the rate of others is notably increased (see Fig. 21).

For the intersubband transition

$$\left\{\left(\nu_i,n_i\right)\rightarrow\left(\nu_f,n_f\right)\&\left(\nu_i,n_j\right)\rightarrow\left(\nu_i,n_g\right)\right\}$$

of type II the resonance condition (41) takes the form

$$\Delta\varepsilon_{if}+\delta\varepsilon_{if}\left(B_{\|}\right)-\frac{\Delta\varepsilon_{if}}{\hbar\omega_{\perp}^{(0)}}\cdot\hbar\omega_{\perp}=0\,. \tag{50}$$

Here, the resonance condition for $B_{\|}=0$ is taken into account:

$$\frac{\Delta\varepsilon_{if}}{\hbar\omega_{\perp}^{(0)}}=\left(n_f-n_i+n_g-n_j\right). \tag{51}$$

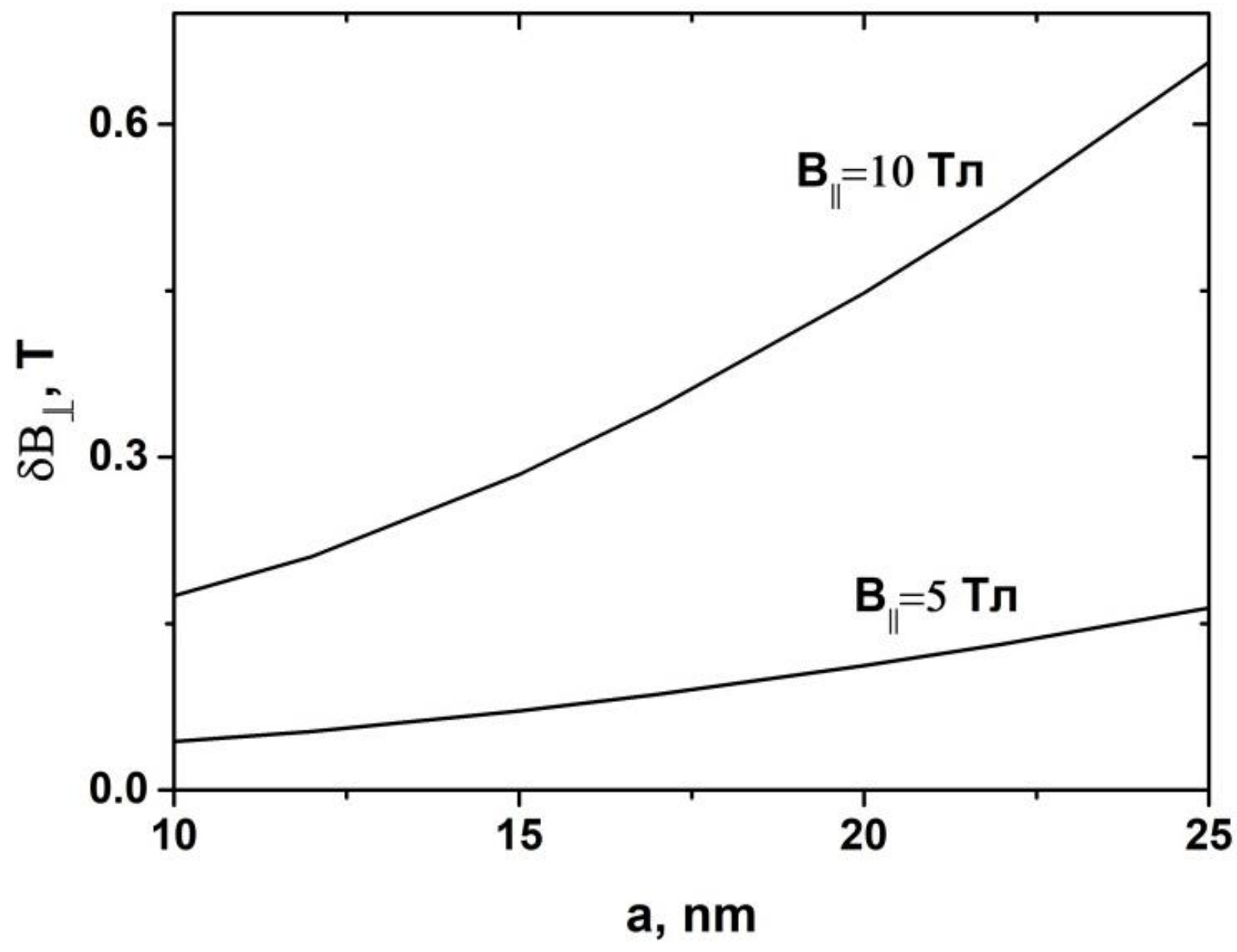

Figure 20. Dependence of the resonance shift of the intersubband transition of type I in tilted magnetic field of the quantum well width. The data are given for the transition $\left\{(2,0)\rightarrow(1,3)\,\&\,(2,0)\rightarrow(1,4)\right\}$.

It follows from (50) that the relations (44) and (45) are also valid for intersubband transitions of type II, from which, taking into account (51), we find

$$\delta B_{\perp} = \frac{mc}{e\hbar} \frac{\delta\varepsilon_{if}\left(B_{\parallel}\right)}{n_f + n_g - n_i - n_j} = \frac{e}{2\hbar c} \frac{\left\lfloor (\delta z)^2_{\nu_i} - (\delta z)^2_{\nu_f} \right\rfloor}{n_f + n_g - n_i - n_j} \cdot B_{\parallel}^2 \tag{52}$$

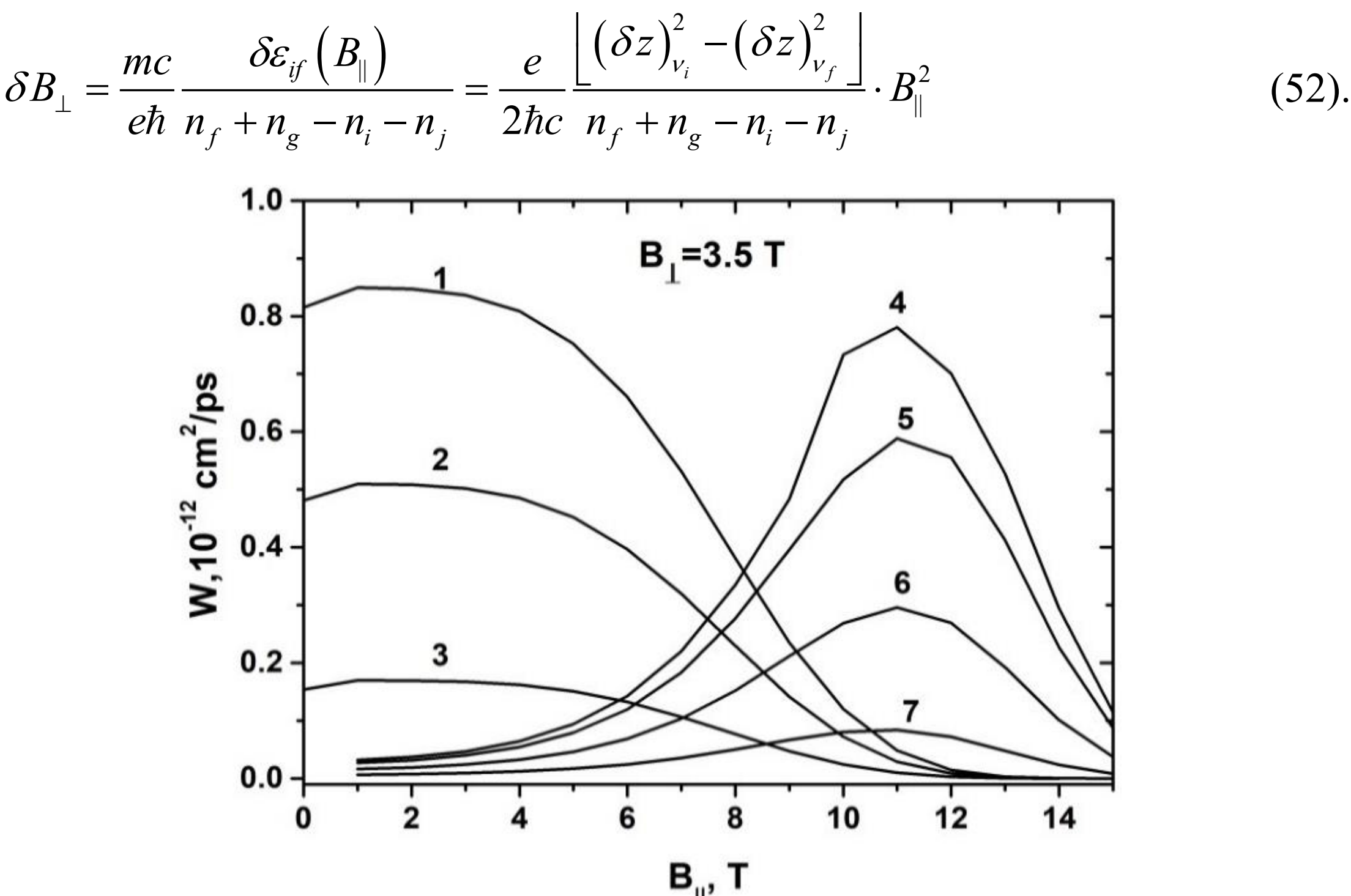

Figure 21. Dependence of the transition rate on $B_{\parallel}$ at the fixed value of $B_{\perp}$=3.5 T.
1 - $\{(2,0)\rightarrow(1,3)\,\&\,(2,0)\rightarrow(1,4)\}$; 2 - $\{(2,0)\rightarrow(1,2)\,\&\,(2,0)\rightarrow(1,5)\}$;
3 - $\{(2,0)\rightarrow(1,1)\,\&\,(2,0)\rightarrow(1,6)\}$; 4 - $\{(2,0)\rightarrow(1,4)\,\&\,(2,0)\rightarrow(1,4)\}$;
5 - $\{(2,0)\rightarrow(1,3)\,\&\,(2,0)\rightarrow(1,5)\}$; 6 - $\{(2,0)\rightarrow(1,2)\,\&\,(2,0)\rightarrow(1,6)\}$;
7 - $\{(2,0)\rightarrow(1,1)\,\&\,(2,0)\rightarrow(1,7)\}$.

The component $B_{\parallel}$ enters expression (9) for the scattering rate amplitude only through the parameter $\xi$, which is determined by expression (18). In electron-electron scattering, two electrons change their states, resulting in two associated parameters in the scattering amplitude: $\xi_{\nu_f,\nu_i}$ corresponding to the transition of the first electron and $\xi_{\nu_g,\nu_j}$ for the second electron. When both parameters are zero ($\xi_{\nu_f,\nu_i} = \xi_{\nu_g,\nu_j} = 0$), the scattering rate is independent on $B_{\parallel}$. This parameter is proportional to the difference between the average coordinates of the electron

$$\langle z \rangle_{\nu} = \int dz \left|\varphi_{\nu}(z)\right|^2 z$$

in its initial and final states. If the average coordinates are the same, the parameter $\xi$ becomes zero.

In the case of intrasubband scattering processes of type I $\nu_i = \nu_j = \nu_g = \nu_f$. Accordingly, $\xi_{\nu_f,\nu_i} = \xi_{\nu_g,\nu_j} = 0$, and hence the scattering rate amplitude of any intrasubband transition of type I is independent of the magnetic field component $B_{\|}$. At the same time, the resonance condition is also independent of $B_{\|}$.

In type II intrasubband scattering processes, the electrons occupy different subbands. However, after scattering, each electron remains in the same subband it initially occupied, i.e., $\nu_f = \nu_i$ and $\nu_g = \nu_j$. Accordingly, $\langle z \rangle_{\nu_f} = \langle z \rangle_{\nu_i}$ and $\langle z \rangle_{\nu_g} = \langle z \rangle_{\nu_j}$. Then $\xi_{\nu_f,\nu_i} = \xi_{\nu_g,\nu_j} = 0$, and the amplitude of the scattering rate of any type II intrasubband transition is also independent on $B_{\|}$. Moreover, the resonance condition for any transition of this type does not depend on the value of $B_{\|}$.

Thus, we conclude that in the case of intrasubband scattering processes, the component of the magnetic field $B_{\|}$ parallel to the structure layers does not affect the electron–electron scattering processes.

The symmetry of the potential profile of the structure $U(z)$ determines the effect of the component $B_{\|}$ on the amplitude of intersubband transitions. In the case of a symmetric potential profile ($U(-z) = U(z)$), the wave functions of the subbands exhibit either even or odd symmetry. Therefore, the average coordinates for all Landau levels across the subbands are identical. Hence, for all transitions $\xi_{\nu_f,\nu_i} = \xi_{\nu_g,\nu_j} = 0$.

Therefore, we conclude that in quantum wells with a symmetric potential profile, the amplitude of transitions of all types is independent on $B_{\|}$. The effect of this component of the magnetic field is manifested only in the shift of the resonances of intersubband transitions of types I and II to higher (transitions from the upper subband to the lower subband) or lower (transitions from the lower subband to the upper subband) values of the quantizing component $B_{\perp}$ of the magnetic field. When the resonances are shifted, the maximum value of the scattering rate changes slightly due to the dependence of the amplitude on $B_{\perp}$ (Fig. 22).

In the case of an asymmetric potential profile, the subband wave functions $\varphi_{\nu}(z)$ are not divided by parity. This means that the average coordinates $\langle z \rangle$ of the initial and final states of one (for type II) or both (types I and II) interacting electrons differ. As a result, one or both parameters $\xi$ would be nonzero, and the transition amplitude $A_{e-e}$ would depend on both $B_{\perp}$, and $B_{\parallel}$.

Asymmetry of the potential profile can be achieved either by applying an external field (e.g., an electric field [3]) or by using an asymmetric design of the quantum well itself [4,5]. From the structure of the expression (9), it is evident that the dependence of the scattering rate amplitude on ξ is nonmonotonic. The change in the potential profile required to change ξ entails a change in the wave functions φ (z) of the quantum confinement levels, which enter into the amplitude not only through the parameter ξ, but also through the integral $R_{\nu_i,\nu_j,\nu_g,\nu_f}(z)$. Therefore, the behavior of the amplitude with variation in $B_{\parallel}$ is quite complex and ambiguous, and it should be studied separately for each structural configuration.

## 6. Conclusion

A "complete" electron–electron scattering rate matrix was calculated, containing all transition types involving the Landau levels of two subbands.

It was found that the rates of intrasubband type II transitions (caused by the interaction of electrons in different subbands) are similar in magnitude and can even exceed the rates of intrasubband type I scattering processes, when both electrons of the interacting pair are at the same Landau level.

It is shown that while intrasubband scattering exhibits a rapid, monotonic decrease in the scattering rate with increasing transferred energy, intersubband scattering processes exhibit an ambiguous dependence on the transferred energy, which is defined by the type and variety of intersubband transitions. The scattering rate can either decrease monotonically or increase monotonically. A nonmonotonic dependence of the scattering rate on the energy transferred to the electron during scattering is also possible.

An expression for the electron–electron scattering velocity in a quantizing magnetic field tilted to the plane of the quantum well layers is obtained. This expression is valid when the cyclotron energy is lower than the quantum confinement energy.

It is demonstrated that, over a wide range of magnetic fields, the component $B_{\parallel}$ of the magnetic field in the plane of the layers has a negligible effect on the intrasubband electron–electron scattering rate.

Two aspects can be distinguished in the influence of this component on intersubband scattering.

The first aspect is the influence of $B_{||}$ on the form factor, which is determined by the energy spectrum. The resonance of type I and type II intersubband transitions shifts to different magnetic field values due to the increase in the intersubband distance caused by the $B_{||}$ component. At the same time, the parallel magnetic field component does not change the resonance condition for type III transitions.

The second aspect is the influence of $B_{||}$ on the intersubband transition amplitude, which does not depend directly on energy but is determined by electron wave functions.

It is shown that the nature of this influence is determined by the symmetry of the quantum well potential profile. In a structure with a symmetric profile, this influence is virtually absent. In the case of an asymmetric potential profile, the amplitude depends on both $B_{\perp}$ and $B_{||}$.